\def\dir{./}
\documentclass[useAMS,usenatbib]{\dir mn2e}
\usepackage[fleqn]{amsmath}
\usepackage{amssymb}
\usepackage[super]{nth}
\usepackage{natbib}
\usepackage{graphicx}
\usepackage{float}
\usepackage{mathrsfs}
\usepackage{mathtools}
\usepackage{multirow}
\usepackage[usenames,dvipsnames]{color}
\usepackage{\dir aas_macros}
\usepackage{comment}
\usepackage{subfigure}
\usepackage{hyperref}
\usepackage{mathrsfs}
\usepackage{epstopdf, dcolumn}
\DeclareGraphicsExtensions{.pdf}
\usepackage{wasysym}
\usepackage{dashrule}
\usepackage{xcolor}
\usepackage{arydshln}
\usepackage{threeparttable}
\usepackage[export]{adjustbox}
\usepackage[figuresleft]{rotating}

\newsavebox{\measurebox}

\newlength\replength
\newcommand\repfrac{.33}

\newcommand\rulewidth{.6pt}
\newcommand\tdashfill[1][\repfrac]{\cleaders\hbox to \replength{%
		\smash{\rule[\arraystretch\ht\strutbox]{\repfrac\replength}{\rulewidth}}}\hfill}

\newcommand\tdotfill[1][\repfrac]{\cleaders\hbox to \replength{%
		\smash{\raisebox{\arraystretch\dimexpr\ht\strutbox-.1ex\relax}{.}}}\hfill}

\newcommand{\appropto}{\mathrel{\vcenter{
			\offinterlineskip\halign{\hfil$##$\cr
				\propto\cr\noalign{\kern2pt}\sim\cr\noalign{\kern-2pt}}}}}

\newcommand{\cmmc}{\textsc{\small 21CMMC}}
\newcommand\lsim{\mathrel{\rlap{\lower4pt\hbox{\hskip1pt$\sim$}}
        \raise1pt\hbox{$<$}}}
\newcommand\gsim{\mathrel{\rlap{\lower4pt\hbox{\hskip1pt$\sim$}}
        \raise1pt\hbox{$>$}}}
\newcommand{\Rom}[1]{\uppercase\expandafter{\romannumeral #1}}
\newcommand{\rom}[1]{\lowercase\expandafter{\romannumeral #1}}

\newcommand{\hone}{\mathrm{H}\textsc{i}}
\newcommand{\hii}{\mathrm{H}\textsc{ii}}
\newcommand{\tocm}{{$\mathrm{21cm}\textsc{fast}$}}
\newcommand{\lya}{{Ly$\alpha$}}
\newcommand{\mvir}{M_{\rm vir}}
\newcommand{\msol}{{\rm M}_\odot}

\defcitealias{Park2019MNRAS.484..933P}{P19}

\begin{document}

\title[Inference from the {\lya} forest]{Reionization and galaxy inference from the high-redshift {\lya} forest}

\author[Qin et al.]{Yuxiang Qin$^{1}$\thanks{E-mail: Yuxiang.L.Qin@gmail.com}, Andrei Mesinger$^{1}$, Sarah E. I. Bosman$^{2,  3}$ and Matteo Viel$^{4,5,6,7}$\\
	$^{1}$Scuola Normale Superiore, Piazza dei Cavalieri 7, I-56126 Pisa, Italy\\
	$^{2}$Department of Physics and Astronomy, University College London, Gower Street, London WC1E 6BT, UK\\
	$^{3}$Max-Planck-Institut f\"{u}r Astronomie, K\"{o}nigstuhl 17, D-69117 Heidelberg, Germany\\
	$^{4}$SISSA - International School for Advanced Studies, via Bonomea 265, 34136 Trieste, Italy\\
    $^{5}$IFPU, Via Beirut 2, 34014, Trieste, Italy\\
    $^{6}$INFN, sezione di Trieste, via Valerio 2, Trieste, Italy\\
    $^{7}$INAF - Osservatorio Astronomico di Trieste, Via G.B. Tiepolo 11, I-34131 Trieste, Italy}
\maketitle
\label{firstpage}

\begin{abstract}
The transmission of Lyman-$\alpha$ ({\lya}) in the spectra of distant quasars depends on the density, temperature, and ionization state of the intergalactic medium (IGM). Therefore, high-redshift ($z>5$)  \lya\ forests could be invaluable in studying the late stages of the epoch of reionization (EoR), as well as properties of the sources that drive it. 
Indeed, high-quality quasar spectra have now firmly established the existence of large-scale opacity fluctuations at $z>5$, whose physical origins are still debated.
Here we introduce a Bayesian framework capable of constraining the EoR and galaxy properties by forward-modelling the high-$z$ {\lya} forest.
Using priors from galaxy and CMB observations, we demonstrate that the final overlap stages of the EoR (when $>95$\% of the volume was ionized) should occur at $z<5.6$, in order to reproduce the large-scale opacity fluctuations seen in forest spectra.
However, it is the combination of patchy reionization and the inhomogeneous UV background that produces the longest Gunn-Peterson troughs.
\lya\ forest observations tighten existing constraints on the characteristic ionizing escape fraction of galaxies, with the combined observations suggesting $f_{\rm esc} \approx 7^{+4}_{-3}$\%, and disfavoring a strong evolution with the galaxy's halo (or stellar) mass.
\end{abstract}


\section{Introduction}
The presence of residual neutral hydrogen in the intergalactic medium (IGM), leads to a series of absorption lines in the spectra of distant galaxies and quasars, corresponding to photons redshifting into Lyman-$\alpha$ (Ly$\alpha$) resonance \citep{Savaglio2002ApJ...567..702S,Lee2014ApJ...795L..12L, FaucherGiguere2008ApJ...681..831F,Busca2013A&A...552A..96B,Lee2013AJ....145...69L,Slosar2013JCAP...04..026S,irsic2017MNRAS.466.4332I}.
This so-called {\lya} forest provides invaluable insights into the structure and properties of the high-redshift IGM (e.g. \citealt{Bolton2010MNRAS.406..612B,Lidz2010ApJ...718..199L,Garzilli2012MNRAS.424.1723G,Lee2015ApJ...799..196L,Puchwein2015MNRAS.450.4081P,Bolton2017MNRAS.464..897B,Gaikwad2020MNRAS.494.5091G}), the cosmic radiation fields that regulate them (e.g. \citealt{Bolton2007MNRAS.382..325B,FaucherGiguere2008ApJ...682L...9F,Becker2013MNRAS.436.1023B,DAloisio2018MNRAS.473..560D}), as well as physical cosmology (e.g. \citealt{MiraldaEscude1996ApJ...471..582M,Croft2002ApJ...581...20C,Viel2005PhRvD..71f3534V,Viel2013PhRvD..88d3502V,Delubac2015A&A...574A..59D,Bautista2017A&A...603A..12B,Yeche2017JCAP...06..047Y}). 

Of particular note is the potential of the {\lya} forest in
studying the observationally-starved epoch of reionization (EoR). Indeed historically, the {\lya} forest provided the first constraint on the ionization state of our Universe \citep{gunn1965density}.

Recent years have witnessed a large increase in the number of high-quality, high-redshift ($z\gsim 5-6$) quasar spectra (e.g. \citealt{Fan2006AJ....132..117F,Becker2007ApJ...662...72B,Willott2010AJ....139..906W,Mortlock2011Natur.474..616M,Venemans2013ApJ...779...24V,Wu2015Natur.518..512W,Becker2015MNRAS.447.3402B,Jiang2016ApJ...833..222J,Banados2018Natur.553..473B,Yang2020ApJ...897L..14Y}). These have been used to search for increasingly subtle EoR signatures.
Redward of the {\lya} emission line, an incomplete EoR can be studied through absorption from the {\lya} damping-wing profile \citep{Bolton2011MNRAS.416L..70B,Schroeder2013MNRAS.428.3058S,Davies2018ApJ...864..142D,Greig2017MNRAS.466.4239G,Greig2019MNRAS.484.5094G,Wang2020ApJ...896...23W},
while on the blueward side, the additional ionizing contribution from the quasar itself facilitates {\lya} transmission in the so-called near zone (e.g.~\citealt{Mesinger2004ApJ...613...23M,Bolton2007MNRAS.374..493B,Lidz2007ApJ...670...39L,Maselli2007MNRAS.376L..34M,Eilers2017ApJ...840...24E,Eilers2020arXiv200201811E,Davies2020MNRAS.493.1330D}).
Blueward of the quasar near zone, the {\lya} forest becomes more opaque and trace amounts of {$\hone$} are sufficient to saturate transmission.

However, some transmission in the forest is seen even at the highest redshifts. In fact, the sightline-to-sightline {\it scatter} in the transmission has been suggested as a potential probe of the EoR (e.g. \citealt{Becker2007ApJ...662...72B,Gallerani2008MNRAS.386..359G,Bosman2018MNRAS.479.1055B,Eilers2018ApJ...864...53E}).
The significant scatter recently observed on large scales (tens - hundred cMpc) is especially promising.  For example, the 110 $h^{-1}$cMpc Gunn-Peterson trough observed in ULAS J0148+0600 cannot be explained by fluctuations in the gas density alone (e.g.~\citealt{Becker2015MNRAS.447.3402B}).

There have been several theoretical explanations for the sizeable sightline-to-sightline fluctuations,  focusing on (i) gas temperature (e.g.~\citealt{DAloisio2015ApJ...813L..38D,Keating2018MNRAS.477.5501K}); (ii) rare sources (\citealt{Chardin2015MNRAS.453.2943C,Chardin2017MNRAS.465.3429C,DAloisio2017MNRAS.468.4691D, Meiksin2020MNRAS.491.4884M}); (iii) the  mean free path of ionizing UV photons \citep{Davies2016MNRAS.460.1328D,DAloisio2018MNRAS.473..560D}; and (iv) late reionization \citep{Kulkarni2019MNRAS.485L..24K,Keating2020MNRAS.491.1736K,Nasir2020MNRAS.494.3080N}. 

Unfortunately, there are significant degeneracies between these models (see e.g.~\citealt{Nasir2020MNRAS.494.3080N}), and most previous work has been fairly qualitative -- showing a handful of models which agree with the observations to various degrees.  Robust, quantitative constraints require an exhaustive Bayesian inference framework. For instance, \citet{Choudhury2020arXiv200308958C} recently used semi-numerical simulations 
to constrain the EoR history using forward-modelled cumulative distribution functions (CDFs) of the effective optical depth $\tau_{\rm eff}$.  Their seminal work, appearing as this paper was nearing completion, used effective parameters to approximate inhomogeneous recombinations, in addition to assuming a constant mass-to-light ratio for galaxies.

Here, we showcase a fully Bayesian framework for interpreting the {\lya} forest at high redshift using {\tocm}\footnote{\url{https://github.com/21cmfast/21cmFAST}} (\citealt{Mesinger2007ApJ...669..663M,Mesinger2011MNRAS.411..955M,Murray2020};) and its Markov chain Monte Carlo (MCMC) driver, {\cmmc}\footnote{\url{https://github.com/21cmfast/21CMMC}} \citep{Greig2015MNRAS.449.4246G,Greig2017MNRAS.472.2651G}. Building on the model introduced by \citet{Park2019MNRAS.484..933P}, we directly sample galaxy properties and forward-model the 3D lightcone of {\lya} transmission. Our galaxy-driven model allows us to fold-in observations of high-redshift UV luminosity functions (LFs; \citealt{Bouwens2015a,Bouwens2016,Oesch2016}), in addition to EoR constrains from the Thomson scattering optical depth ($\tau_e$; \citealt{Planck2018arXiv180706209P}) and quasar dark fraction measurements \citep{McGreer2015MNRAS.447..499M}. All codes developed here are publicly available.

This paper is organized as follows. We present our model including galaxy properties, reionization, IGM temperature and {\lya} forests in Section \ref{sec:models}. We then summarize the observed forest sample and free parameters used for inference in Sections \ref{sec:observation} and \ref{sec:parameters}. We present our inference results in Section \ref{sec:inference} before concluding in Section \ref{sec:conclusions}. We assume a $\Lambda$CDM cosmology with parameters
($\Omega_{\mathrm{m}}, \Omega_{\mathrm{b}}, \Omega_{\mathrm{\Lambda}}, h, \sigma_8, n_s $ = 0.31, 0.049, 0.69, 0.68, 0.81, 0.97) chosen from the {\it TT,TE,EE+lowE+lensing+BAO} reconstruction in \citet{Planck2018arXiv180706209P}.

\section{Modelling galaxies, the IGM and {\lya} transmission}\label{sec:models}

We first summarize our parametrization of high-redshift galaxies, whose cosmic radiation fields govern the evolution of the high-redshift IGM.  We then discuss the corresponding IGM properties, including inhomogenous reionization, local recombination and photoionization rates, before presenting the addition of two new outputs of {\tocm} developed in this work: post-reionization gas temperature and the {\lya} optical depth.

\subsection{High-redshift galaxies}\label{subsec:21cmfast}

We adopt the galaxy model of \citet{Park2019MNRAS.484..933P}, which relates bulk galaxy properties to the halo mass function through power-law scalings.  Such a parametrization can recover high-redshift galaxy UV LFs (see also, e.g.~\citealt{Moster2013MNRAS.428.3121M,Sun2015,Mutch2016,Tacchella2018ApJ...868...92T,Behroozi2019MNRAS.488.3143B,Yung2019MNRAS.490.2855Y}).  Therefore, we can use galaxy observations, in addition to the {\lya} forest, to constrain our model parameters. This improves on some previous forest studies (e.g.~\citealt{Mesinger2009MNRAS.400.1461M, Choudhury2020arXiv200308958C}) that assumed a constant mass to light ratio,
and allows us to use physically-meaningful priors when performing inference (e.g.~ionizing escape and stellar mass fractions must be between 0 and 1).

Specifically, the number density of galaxies is described by the halo mass function, ${\rm d}n/{\rm d}{\mvir}$,
with an additional factor of $\exp\left(-{M_{\rm turn}}/{\mvir}\right)$ accounting for inefficient star formation in low-mass halos due to ineffective cooling, inhomogeneous feedback from reionization photo-heating and/or supernova feedback (\citealt{Efstathiou1992MNRAS.256P..43E,Shapiro1994ApJ...427...25S,Thoul1996ApJ...465..608T,Hui1997MNRAS.292...27H,Sobacchi2014MNRAS.440.1662S,Hopkins2014MNRAS.445..581H, Wyithe2013MNRAS.428.2741W,Sun2015,Mutch2016,Hopkins2017}).
The average stellar mass of a galaxy hosted by a halo of mass $\mvir$ can be written as
$M_*{=}\mvir\Omega_{\mathrm{b}}/\Omega_{\mathrm{m}}\times\min\left[1, f_{*,10} \left({M_{\rm vir}}/{10^{10}{\msol}}\right)^{\alpha_*}\right]$.  Similarly, the UV ionizing escape fraction is taken to be
$f_{\rm esc}{=}\min\left[1, f_{\rm esc,10} \left({M_{\rm vir}}/{10^{10}{\msol}}\right)^{\alpha_{\rm esc}}\right]$. Assuming the characteristic star formation timescale is proportional to the halo dynamical timescale, we take $M_*/\dot{M}_*{=}t_*H^{-1}$ where $H(z)$ represents the Hubble parameter at $z$.  We can then estimate the non-ionizing UV luminosity through $L_{1500} {=} \dot{M}_*{\times}8.7{\times}10^{27} {\rm erg\ s^{-1}Hz^{-1}} \msol^{-1} {\rm yr}$ \citep{Madau2014ARA&A..52..415M} and compare against high-redshift observations (e.g.~\citealt{Finkelstein2015ApJ...810...71F,Bouwens2015a,Bouwens2016,Livermore2017ApJ...835..113L,Atek2018MNRAS.479.5184A,Ishigaki2018ApJ...854...73I,Oesch2018ApJ...855..105O,Bhatawdekar2019MNRAS.tmp..843B}).

We thus have $6$ free parameters to characterize the UV ionizing properties of high-redshift galaxies: $M_{\rm turn}$, $f_{*,10}$, $\alpha_*$, $f_{\rm esc,10}$, $\alpha_{\rm esc}$ and $t_*$. In the next subsection, we summarize how we calculate the IGM properties corresponding to a given galaxy model.

\subsection{The IGM}\label{subsec:temp}

We begin by generating a Gaussian realization of $\Lambda$CDM initial conditions in a periodic box with a side length of $500$cMpc and a cell resolution of ${\sim}0.39$cMpc (i.e. $500$cMpc$/1280$). Using second-order Lagrangian perturbation theory \citep{Scoccimarro1998MNRAS.299.1097S} with high-res velocity and density fields\footnote{Rather than evolving the density field using lower resolution velocity fields (which was originally implemented in {\tocm} to conserve RAM, given that velocity fields have much longer correlation lengths), here we use high-resolution for all of the initial conditions (setting \textsc{perturb\_on\_high\_res=true}). This was included to guarantee the density fields are as accurate as possible for simulating the {\lya} forest (Watkinson et al., in prep).}, we evolve these fields towards lower redshifts and re-grid them onto a lower resolution box (${\sim}1.95$cMpc; i.e.~$500$cMpc$/256$) to calculate ionization fields.

\subsubsection{Inhomogeneous reionization}
Cosmological reionization by UV photons is effectively bi-modal, with ionized regions surrounding galaxies expanding and eventually overlapping to complete reionization.  EoR models need to track this inhomogenous process, as well as estimate the residual $\hone$ inside the ionized component of the IGM.  We summarize our procedures for these in turn.

To identify ionized cells, we use an excursion-set approach\footnote{Excursion-set algorithms generally do not conserve photons when $\hii$ bubbles overlap (e.g. \citealt{Zahn2007ApJ...654...12Z,Paranjape2014MNRAS.442.1470P}).
In practice, this translates to a bias in the effective ionizing escape fraction (e.g.~\citealt{Hutter2018MNRAS.477.1549H}).
Using the updated, photon-conserving algorithm of {\tocm} v3,
 we quantify that this is a very minor effect for our model parametrization, resulting in a bias of ${\sim}$ -0.2 for $\alpha_{\rm esc}$ (Park et al., in prep).  In other words, the recovered posterior without ionizing photon conservation differs from the true posterior including ionizing photon conservation primarily through a translation in one of the parameters: $\alpha_{\rm esc}^{\rm true} \rightarrow \alpha_{\rm esc}^{\rm recovered} + 0.2$.  However, as photon-conservation slows down our calculation by a factor of $\sim$2, we leave this option off in this proof-of-concept study, highlighting the resulting bias in the marginalized posterior of $\alpha_{\rm esc}$ (see more in Section \ref{sec:inference}).} \citep{Furlanetto2004ApJ...613....1F}. Centered on a cell at (${\bf r},z$), we consider spherical volumes with decreasing radii, $R$, and corresponding overdensities, $\delta_{\rm R|_{{\bf r},z}}{\equiv} \langle \rho_{\rm b}/\bar{\rho}_{\rm b} {-}1 \rangle_R$ where $\rho_{\rm b}$ and $\bar{\rho}_{\rm b}$ are the baryon density and its cosmic mean. Using the corresponding conditional halo mass function \citep{Barkana2005ApJ...626....1B, Mesinger2011MNRAS.411..955M},
 we compute the cumulative number of ionizing photons per baryon in this spherical IGM patch by
\begin{equation}\label{eq:n_ion}
	\bar{n}_{\rm ion} =  \int {\rm d}M_{\rm vir} \frac{{\rm d}n}{{\rm d}{\mvir}}\exp\left({-}\frac{M_{\rm turn}}{\mvir}\right) {M_*}{\rho_{\rm b}^{-1}}n_{\gamma} f_{\rm esc},
\end{equation}
where $n_{\gamma}{=}5000$ is the number of ionizing photons intrinsically emitted per stellar baryon \citep{Barkana2005ApJ...626....1B}. 

We follow \citet{Sobacchi2014MNRAS.440.1662S} to estimate the average number of recombinations per baryon, $\bar{n}_{\rm rec}$. Using the probability distribution function (PDF; ${\rm d}n/{\rm d}\rho_{\rm sub}$) of sub-grid (unresolved by our simulation cell ${\lesssim}1.95$cMpc) densities ($\rho_{\rm sub}$) from \citet{Miralda2000ApJ...530....1M}, adjusted for the mean density in the cell, 
we calculate the recombination rate by
\begin{equation}\label{eq:n_rec}
\dot{n}_{\rm rec} =\int{\rm d}\rho_{\rm sub}\frac{{\rm d}n}{{\rm d}\rho_{\rm sub}}\alpha_{\rm B}f_{\rm H} {\rho_{\rm b}^{-1}}{\rho_{\rm sub}^2} \left(1{-}x_{\rm \hone}\right)^2,
\end{equation}
where $\alpha_{\rm B}$, $f_{\rm H}$ and $x_{\hone}$ are the case-B recombination coefficient\footnote{The recombination coefficient depends on the gas temperature ($T_{\rm g}$), $\alpha_{\rm B}{=}2.59{\times}10^{-10} (T_{\rm g}/{\rm K})^{-0.75}{\rm cm^3 s^{-1}}$. For computational efficiency, we assume an average temperature of $T_{\rm g}{=}10^4{\rm K}$ in equation (\ref{eq:n_rec}) when computing the {\it cumulative} number of recombinations, used to identify if a region is ionized or not.  We do account for local temperature fluctuations when computing the residual {$\hone$} fraction inside the ionized IGM, as detailed in Section \ref{sec:residualHI}.}, number fraction of hydrogen in the Universe, and the (residual) neutral hydrogen fraction of the sub-grid gas element, respectively. The calculation of $x_{\hone}$ is presented in Section \ref{sec:residualHI}.

The cell is then considered to be ionized if the cumulative number of ionizing photons is larger than the number of baryons plus recombinations.  Specifically,
the cell is considered as ionized if at any radius, $R$, 
\begin{equation}\label{eq:ionization}
\bar{n}_{\rm ion} - 1 \ge \bar{n}_{\rm rec} = \langle{\int_{z_{\rm ion}}^{z} {\rm d}z^\prime\frac{{\rm d}t}{{\rm d}z^\prime} \dot{n}_{\rm rec}} \rangle_{R},
\end{equation}
where $z_{\rm ion}$ is the reionization redshift of a cell and $\langle\rangle_{R}$ denotes averaging over all cells within the spherical $\hii$ region. We also approximate the local mean free path ($R_{\rm MFP}$) using the largest radius at which this equation is valid. This  is strictly true in the early stages of reionization that are not affected by IGM recombinations, but should also be a good approximation for the overlap stages as they likely evolve in a ``photon-starved" manner  \citep{Bolton2007MNRAS.382..325B, Sobacchi2014MNRAS.440.1662S}. 
In future work, we will generalize this derivation, which will allow us to extend our models to lower redshifts.

\subsubsection{Residual neutral hydrogen inside the ionized IGM}
\label{sec:residualHI}

Since a trace amount of neutral hydrogen
can obscure all flux at the {\lya} transition, it is important to determine the residual neutral hydrogen fraction within the cosmic $\hii$ regions. Assuming photoionization 
equilibrium in the reionized IGM, we evaluate $x_{\rm \hone}$ by solving
\begin{equation}\label{eq:photoionization_equilibrium}
x_{\hone} f_{\rm ion,ss}\Gamma_{\rm ion} = \chi_{\rm HeII} n_{\rm H} (1-x_{\hone})^2\alpha_{\rm B}
\end{equation}
where $\Gamma_{\rm ion}$ is the local photoionization background\footnote{Our model does not include small-scale fluctuations in $\Gamma_{\rm ion}$, due to proximate galaxies.  In most of the IGM, such Poisson fluctuations are negligible, and $\Gamma_{\rm ion}$ is determined by the combined radiation from many galaxies (e.g. \citealt{Mesinger2009MNRAS.400.1461M,Sadoun2017ApJ...839...44S}).}, $f_{\rm ion,ss}$ is a self-shielding factor attenuating $\Gamma_{\rm ion}$, $\chi_{\rm HeII}{=}1.08$ accounts for singly ionized helium, and $n_{\rm H}$ is the hydrogen number density in the cell. 

The photoionization rates before self-shielding are estimated assuming a stellar-driven UV background (UVB) with a spectral index\footnote{The effective spectral index for a stellar-driven UVB could be somewhat harder than we assume (e.g. \citealt{Becker2013MNRAS.436.1023B,DAloisio2019ApJ...874..154D}); however, $\alpha_{\rm UVB}$ is degenerate with the ionizing escape fraction in equation \ref{eq:gamma}, and we treat the later as a free parameter in our analysis.} 
of $\alpha_{\rm UVB}{\sim}5$ \citep{Thoul1996ApJ...465..608T}
\begin{equation}\label{eq:gamma}
{\Gamma}_{\rm ion} = \left(1+z\right)^2 R_{\rm MFP}\sigma_{\rm H}\frac{\alpha_{\rm UVB}}{\alpha_{\rm UVB}+\beta_{\rm H}}\frac{\bar{\rho}_{\rm b}}{m_{\rm p}}{\bar{\dot{n}}_{\rm ion}},
\end{equation}
where $\beta_{\rm H}{\sim}2.75$, $m_{\rm p}$ and 
$\bar{\dot{n}}_{\rm ion}$ are the spectral index of the {$\hone$} photoionization cross-section, 
the proton mass, and the mean production rate of ionizing photons evaluated using equation (\ref{eq:n_ion}) with 
$M_*$ being replaced by $\dot{M}_*$. 
According to radiative transfer 
simulation results from \citet{Rahmati2013MNRAS.430.2427R}, the self-shielding factor depends on the local density ($\rho_{\rm b}$), gas temperature ($T_{\rm g}$) as well as the unattenuated photoionization rate (${\Gamma}_{\rm ion}$) and follows
\begin{equation}
{f_{\rm ion,ss}} = 0.98\left[1{+}\left(\frac{\rho_{\rm b}}{\rho_{\rm ss}}\right)^{1.64}\right]^{-2.28} {+} 0.02\left(1{+}\frac{\rho_{\rm b}}{\rho_{\rm ss}}\right)^{-0.84} 
\end{equation}
where
\begin{equation}
{\rho_{\rm ss}} = 27\bar{\rho}_{\rm b}\left(\frac{T_{\rm g}}{10^4{\rm K}}\right)^{0.17}\left(\frac{\Gamma_{\rm ion}}{10^{-12}{\rm s}^{-1}}\right)^{\frac{2}{3}}\left(\frac{1+z}{10}\right)^{-3}
\end{equation}
is the characteristic density for the onset of self-shielding
\citep{Schaye2001ApJ...559..507S}.

\subsubsection{Gas temperature of the ionized IGM}
Inside the {$\hii$} regions, we calculate the inhomogeneous gas temperature in each simulation cell following \citet{McQuinn2016MNRAS.456...47M}:
\begin{equation}\label{eq:tg}
T_{\rm g}^{\gamma} = T_{\rm ion,I}^{\gamma} \left[\left(\frac{\mathscr{Z}}{\mathscr{Z}_{\rm ion}}\right)^{3} \frac{\rho_{\rm b}}{\rho_{\rm b,ion}}\right]^{\frac{2\gamma}{3}} \frac{\exp\left(\mathscr{Z}^{2.5}\right)}{\exp\left(\mathscr{Z}_{\rm ion}^{2.5}\right)} + T_{\rm lim}^{\gamma}\frac{\rho_{\rm b}}{\bar{\rho}_{\rm b}}
\end{equation}
Here, $\mathscr{Z}$ denotes $(1+z)/7.1$,
$\gamma {=} 1.7$ is the equation of state index, $T_{\rm ion,I}$ represents the post I-front temperature, here taken to be a constant ($T_{\rm ion,I} = 2\times10^{4}{\rm K}$)\footnote{The exact value depends on I-front speeds, and to a more minor extent, the spectral index of the ionizing background.  However,  the uncertainty and scatter in this value should be smaller than the scatter resulting from different reionization times (e.g. \citealt{DAloisio2019ApJ...874..154D, Davies2019MNRAS.489..977D,Zeng2020arXiv200702940Z}).}, and $T_{\rm lim}=1.775\mathscr{Z}\times10^{4}{\rm K}$ refers to the final relaxation temperature.
The subscript ``$_{\rm ion}$'' indicates the quantity is at the redshift of ionization, $z_{\rm ion}$. The two terms on the RHS of equation (\ref{eq:tg}) correspond to the initial and final temperatures of photo-ionized IGM.

The memory of the initial, post-reionization temperature fades within $\Delta z{\sim} 1-2$ (see the exponential term in equation \ref{eq:tg}), after which the IGM approaches an equilibrium temperature resulting from the balance between photoheating and various cooling processes (i.e.~Hubble expansion, recombination, Compton scattering and free-free radiation).  Because reionization is ``inside-out'' on large scales (e.g.~see the review of \citealt{Trac2011ASL.....4..228T}), the underdense regions of the IGM reionize late.  Thus, large-scale low-density regions at $z{\sim}$5--6 should be {\it hotter} (due to the first term on the RHS of equation \ref{eq:tg}), resulting in a lower recombination rate coefficient in equation (\ref{eq:photoionization_equilibrium}) and a corresponding {\it increase} in the {\lya} transmission in the forest.  On the other hand, the equilibrium temperature results in a temperature-density relation in which underdense regions are {\it colder} (see the second term on the RHS of quation \ref{eq:tg}; see also \citealt{Hui1997MNRAS.292...27H}), and have a correspondingly {\it lower} {\lya} transmission.  However, when averaged over large scales, only the former effect remains. Therefore, one would expect voids to correspond to large-scale peaks in the {\lya} forest transmission (e.g.~\citealt{DAloisio2015ApJ...813L..38D}). This correlation between gas temperature and {\lya} transmission is, however, opposite of what is inferred from ULAS J0148+0600 based on galaxy counts \citep{Becker2018ApJ...863...92B,Kashino2020ApJ...888....6K}, suggesting that opacity fluctuations are not dominated by the temperature field for this one sightline.

\subsection{Building physical intuition from examples of forward-modelled data}\label{subsec:forest}
Using the IGM properties described in the previous sections, we compute the corresponding {\lya} optical depth in each simulation cell.  As we are interested in large-scale effects and use relatively low resolution simulations,
we adopt the fluctuating Gunn-Peterson approximation (FGPA\footnote{In Appendix \ref{app:sec:FGPA}, we use high-resolution hydrodynamic simulations to quantify the error in modelling the large-scale {\lya} transition caused by the FGPA.  We include the resulting error covariance in our likelihood calculation, as detailed below.}; \citealt{gunn1965density,Rauch1998ARA&A..36..267R,Weinberg1999elss.conf..346W}) and estimate the optical depth by
\begin{equation}\label{eq:tau_lya}
\begin{split}
\tau_{\alpha} &= f_{\rm rescale} \times \sqrt{\frac{3\pi\sigma_{\rm T}}{8}} c f_{\alpha} \lambda_\alpha H^{-1} n_{\rm H} x_{\hone}
\end{split}
\end{equation}
where $\sigma_{\rm T}$, $f_\alpha{=}0.416$ and $\lambda_\alpha{=}1216\mathrm{\AA}$ are the Thomson cross-section, oscillator strength,
and {\lya} rest-frame wavelength.

To better match observations, numerical simulations usually re-scale the optical depth by a constant factor within a given redshift window (either implicitly or explicitly; e.g. \citealt{Chardin2017MNRAS.465.3429C,DAloisio2018MNRAS.473..560D,Keating2018MNRAS.477.5501K,Bosman2018MNRAS.479.1055B}),
such that the modelled mean {\lya} transmission agrees with the measured values: $\tau_{\alpha}(\lambda_{\rm obs}) \rightarrow \tau_{\alpha} (\lambda_{\rm obs}) \times f_{\rm rescale}$.
This adds additional flexibility to the modelling (e.g. allowing for an arbitrary normalization and redshift evolution of the UV photoionization rate), ameliorates errors in the continuum subtraction, and/or compensates for modelling errors.  However, such re-scaling wastes the predictive power of the model's emissivity and its redshift evolution.
We further explain how we implement this rescaling factor in Section \ref{sec:parameters}.

Below we showcase our procedure for generating mock {\lya} forest data.  As a specific illustrative example, we use the maximum likelihood model in the {\it forest} posterior (presented in Section \ref{sec:inference}).  This model is able to reproduce observations of the high-redshift galaxy UV LFs \citep{Bouwens2015a,Oesch2018ApJ...855..105O}, the CMB optical depth \citep{Planck2018arXiv180706209P}, and the Ly$\alpha$ forest \citep{Bosman2018MNRAS.479.1055B}.
It corresponds to the following astrophysical model parameters -- ($f_{*,10}$, $\alpha_*$, $f_{\rm esc,10}$, $\alpha_{\rm esc}$, $M_{\rm turn}$, $t_*$) = (0.0448, 0.488, 0.0914, -0.298, $7.16\times10^8{\msol}$, 0.378) as well as $f_{\rm rescale}=0.9+0.2\times(z-5.7)$ for flux normalization (see more in Section \ref{sec:parameters}).

\subsubsection{Lightcones and cross correlation coefficients}

\begin{figure}
	\begin{minipage}{\columnwidth}
		\hspace*{-4mm}
		\includegraphics[width=1.03\textwidth]{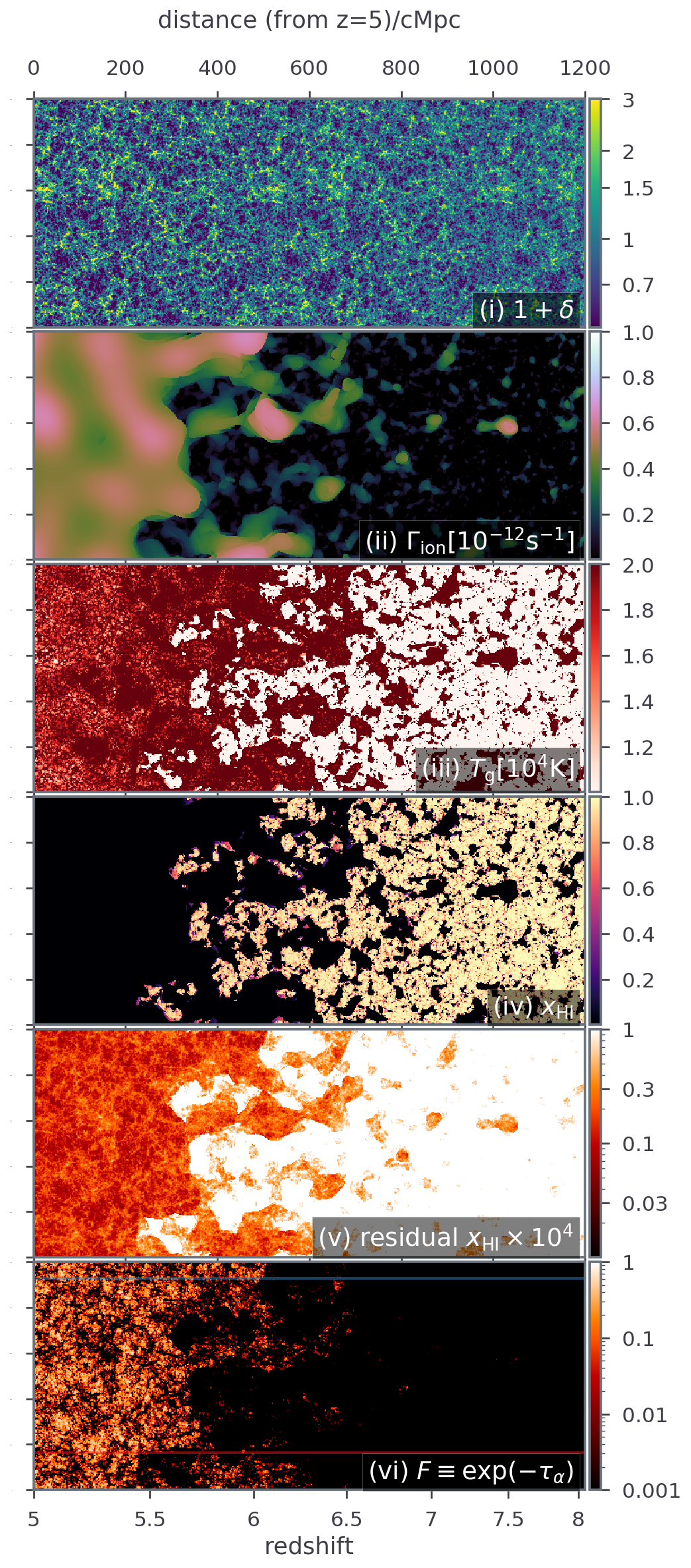}\\
		\vspace*{-5mm}
		\caption{\label{fig:reference}
			2D slices through lightcones with a spatial scale of $500$cMpc on the vertical axis and a thickness of $1.95$ Mpc, spanning a redshift range of $5{<}z{\lesssim}8$ (corresponding to a $1.2$cGpc sight length) for the reference model (i.e.~the maximum likelihood model in the {\it forest} posterior in Fig.~\ref{fig:MCMC}). From top to bottom, the panels correspond to:
			(i) overdensity ($1+\delta$);
			(ii) locally averaged UVB (${\Gamma}_{\rm ion}$ in units of $10^{-12}\rm{s}^{-1}$);
			(iii) temperature ($T_{\rm g}$); 
			(iv) neutral hydrogen fraction ($x_{\hone}$ on a linear scale between $0$ and $1$);
			(v) residual neutral hydrogen fraction ($x_{\hone}$ on a logarithmic scale between $10^{-6}$ and $10^{-4}$); and
			(vi) {\lya} transmission ($F{\equiv}e^{-\tau_{\alpha}}$).  The red and blue lines in the bottom panel mark the two sample sightlines shown in Fig.~\ref{fig:reference_los}.
		}
	\end{minipage}
\end{figure}

\begin{figure}
	\begin{minipage}{\columnwidth}
		\centering
		\includegraphics[width=\textwidth]{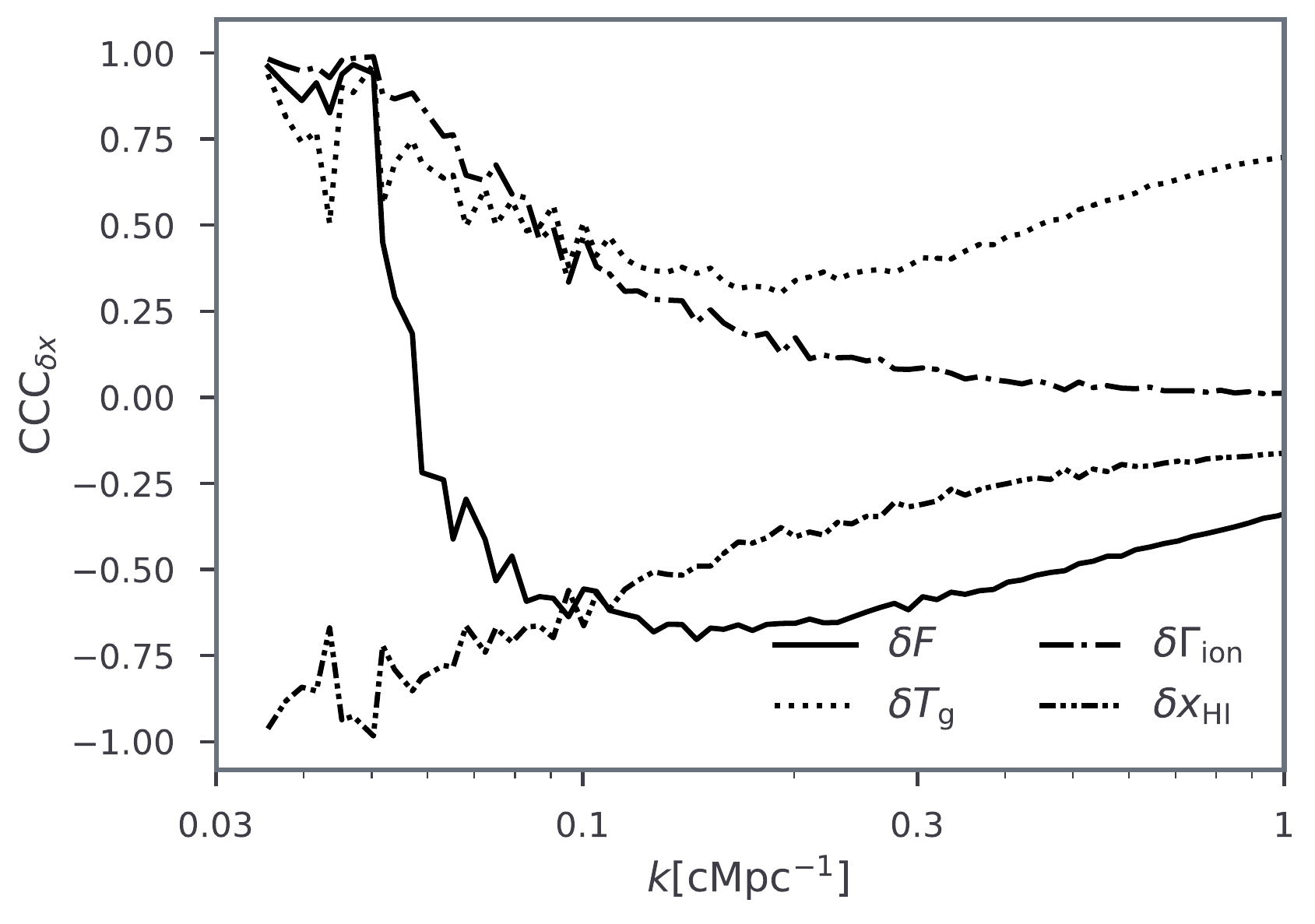}
		\vspace*{-2mm}
		\caption{\label{fig:varying_CCC} Cross correlation coefficients between the overdensity field ($\delta$) and: (i) temperature ($T_{\rm g}$), (ii) photonionization rate ($\Gamma_{\rm ion}$), (iii) neutral hydrogen fraction ($x_{\hone}$), (iv) {\lya} transmission ($F$),  for the reference model at $z=5.8$ ($\bar{x}_{\hone}{\sim}0.14$).}
	\end{minipage}
\end{figure}

In Fig. \ref{fig:reference}, we show lightcone slices (with a thickness of 1.95 Mpc) through the following fields: (i) density ($1+\delta$), (ii)  large-scale photoionization rate ($\Gamma_{\rm ion}$), (iii) gas temperature ($T_{\rm g}$), (iv) the order unity fluctuations in the neutral hydrogen fraction from the EoR ($x_{\hone}$), (v) the residual neutral hydrogen fraction inside the ionized IGM, and (vi) the {\lya} transmission [$F\equiv\exp(-\tau_{\alpha})$]. In addition, to quantify how these fields correlate with the underlying density on various scales, we calculate the cross correlation coefficient (CCC) defined as the cross-correlation power between density and a field, $P_{\rm \delta x}$, normalized by their auto-correlation power: ${\rm CCC}_{\rm \delta x}(k) = {P_{\rm \delta x}(k)}/({\sqrt{P_{\rm \delta \delta}(k)}\sqrt{P_{\rm x x}(k)}})$. Fig. \ref{fig:varying_CCC} shows CCCs between $F$, $T_{\rm g}$, $\Gamma_{\rm ion}$, $x_{\hone}$ and $\delta$ using a snapshot at $z{\sim}5.8$ (see projections in Fig. \ref{fig:tau_alpha_comparisons}). We summarize some key trends below.

\begin{figure*}
	\begin{minipage}{\textwidth}
		\centering
		\includegraphics[width=0.98\textwidth]{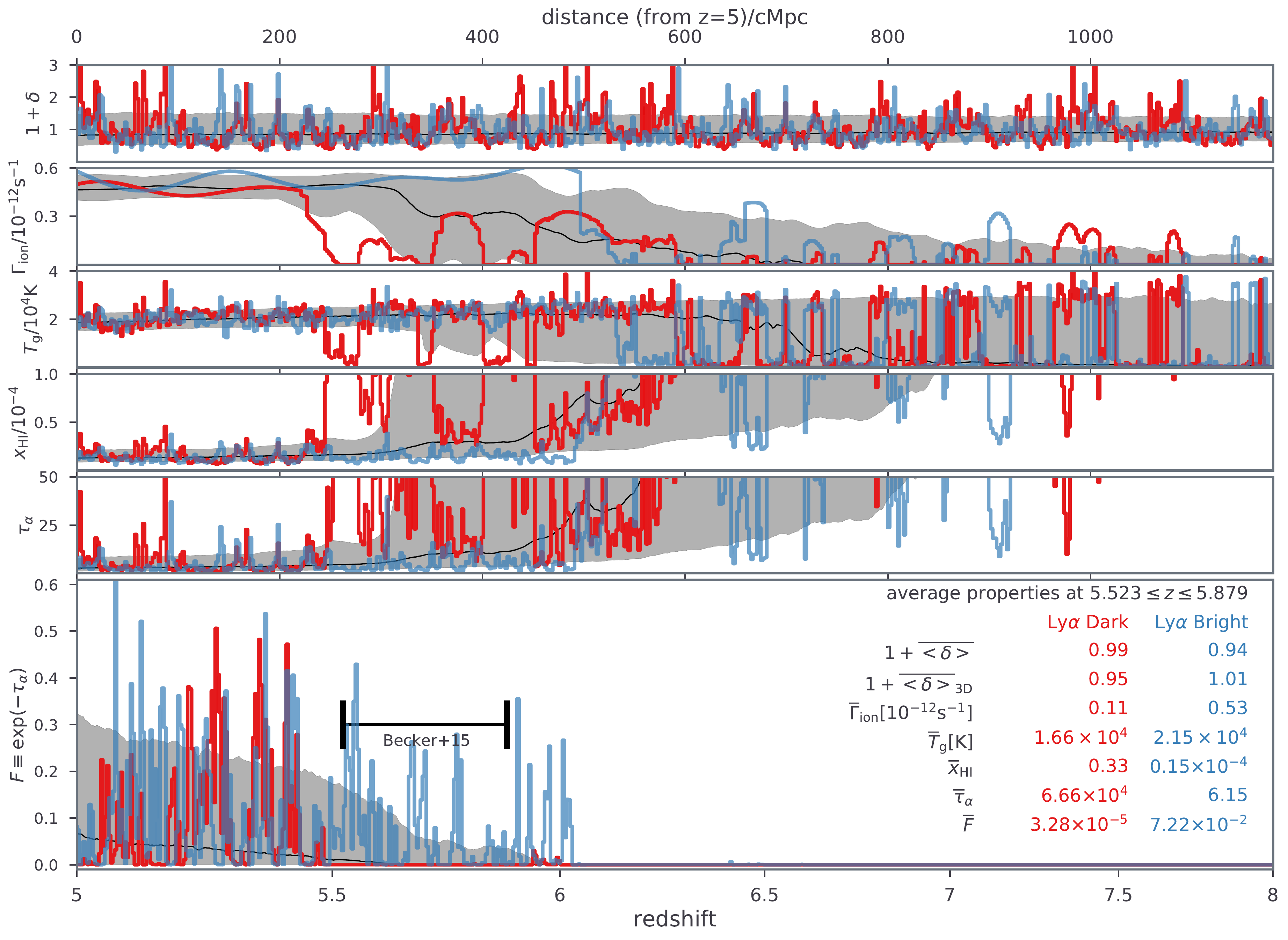}\vspace*{-2mm}
		\caption{\label{fig:reference_los} {\it From top to bottom:} overdensity ($1+\delta$), locally averaged UVB (${\Gamma}_{\rm ion}$), gas temperature ($T_{\rm g}$), residual neutral hydrogen fraction ($x_{\hone}$),  {\lya} optical depth ($\tau_{\alpha}$) and transmission ($e^{-\tau_{\alpha}}$), along two particular lines of sight (i.e. {\it {\lya} Dark} in red and {\it {\lya} Bright} in blue; these two sightlines are also marked by the two horizontal lines in panel (v) of Fig.~\ref{fig:reference}). The sightlines are shown with thick colored curves while the median and [$14,86$] percentiles of the entire 3D lightcone are presented as the thin black lines and shaded regions. The average properties along the sightline at $5.523{\le} z {\le} 5.879$ (the redshift range of the $110h^{-1}$cMpc {\lya} GP trough from \citealt{Becker2015MNRAS.447.3402B}) are listed in the bottom right corner. We also show the density averaged over a 3D volume corresponding to $50{\times}50{\times}110 h^{-3}{\rm cMpc}^{3}$, highlighting that the large-scale environment of {\it {\lya} Dark} is less dense than that of {\it {\lya} Bright}; see text for more details.}
	\end{minipage}
\end{figure*}

\begin{enumerate}
	\item The underlying density (panel i) plays a key role in determining the UV ionizing background (panel ii) and the reionization morphology (panel iv). Regions near high-density peaks become ionized first with the $\hii$ bubbles spreading into the large-scale voids at later times (i.e.~``inside-out" reionization; \citealt{Furlanetto2004ApJ...613....1F,Iliev2006MNRAS.369.1625I,Lee2008ApJ...675....8L,Choudhury2009MNRAS.394..960C,Friedrich2011MNRAS.413.1353F,Bauer2015MNRAS.453.3593B,Mesinger2016ASSL..423.....M,Hutter2017ApJ...836..176H}) and overlapping with each other to complete the EoR (at $z\sim5.5$ in this model). This can also be seen in the CCCs: $x_{\hone}$ anti-correlates with $\delta$ and this anti-correlation strengthens towards smaller $k$ (larger scales).  A similar trend is also found for the photoionization rate which  correlates with density.  The characteristic scales of these correlations depend on the model (i.e.~the luminosity-weighted galaxy bias) and stage of EoR (e.g.~\citealt{McQuinn2007MNRAS.377.1043M, Zahn2011MNRAS.414..727Z}).
	\item The IGM temperature (panel iii) shows both large-scale and small-scale structure.  Large-scale hot regions are evident in the temperature maps down to $z\sim5.2$, after the end of the EoR in this model. These trace the large-scale underdensities that were the last to reionize (panel iv). However, large-scale structure is not evident at lower redshifts, with the IGM cooling towards $T_{\rm lim}(1+\delta)^{1/\gamma}$ (see equation \ref{eq:tg}). Indeed, from Fig. \ref{fig:varying_CCC} we confirm that ${\rm CCC}_{\delta T_{\rm g}}$ at $z=5.8$ has a non-monotonic relation with $k$, which rises towards scales smaller than $k{\sim}0.2{\rm cMpc}^{-1}$.  Smaller scales are dominated by the temperature-density relation, while larger scales are dominated by EoR morphology.
   	\item The residual neutral hydrogen fraction (panel v) has a complex, multi-scale structure as it depends on the density, ionizing background and temperature (see equation \ref{eq:photoionization_equilibrium}). On large-scales ($\gtrsim50$cMpc), the residual $x_{\hone}$ anti-correlates with the $\Gamma_{\rm ion}$ field, while on small-scales it can be seen to correlate/anti-correlate with the density/temperature fields (see also \citealt{DAloisio2015ApJ...813L..38D,Davies2016MNRAS.460.1328D,DAloisio2018MNRAS.473..560D,Keating2018MNRAS.477.5501K}).
	\item The {\lya} optical depth (see the transmission in panel vi) generally  decreases towards lower redshifts with fluctuations following the density field (panel i) and the inferred residual $\hone$ fraction (panel v; see equation \ref{eq:tau_lya}). We can clearly see long patches with {\lya} transmission lower than a thousandth on scales larger than $10$ cMpc post reionization with some even surviving at $z\sim5$. We see that the transmitted flux and density anti-correlate on scales smaller than $k\gtrsim 0.05{\rm cMpc}^{-1}$ (i.e.~${\lesssim}130$cMpc), with the strongest anti-correlation occurring around $k{\sim}0.1{\rm cMpc}^{-1}$ (i.e.~${\sim}60$cMpc). On extremely large scales however {\lya} transmission is tightly correlated with the underlying density through the ionizing background. 
\end{enumerate}

\subsubsection{Spectra}

In Fig.~\ref{fig:reference_los}, we show two example 1D sightlines from this lightcone, denoted by the red ({\it {\lya} Dark}) and blue lines ({\it {\lya} Bright}) in panel (v) of Fig.~\ref{fig:reference}.  {\it {\lya} Dark} was chosen as it exhibits a long, $110h^{-1}$cMpc GP trough over the redshift interval $5.52{\le} z {\le} 5.88$, consistent with ULAS J0148+0600 (\citealt{Becker2015MNRAS.447.3402B}; see the bottom panel of Fig.~\ref{fig:reference_los}).
{\it {\lya} Dark} shows an average transmission of ${<}10^{-4}$ over this redshift range, which is much lower than {\it {\lya} Bright} ($0.072$).  The latter corresponds to a more typical sightline at these redshifts\footnote{Among all $256^2 = 65536$ lines of sight in this maximum likelihood model, we find only $179$ with a mean transmission lower than $0.001$. This is broadly consistent with current observations ($1$ out of ${\sim}300$; \citealt{Bosman2020zndo...3634964B}).}.

\begin{figure*}
	\begin{minipage}{0.44\textwidth}
		\includegraphics[width=\textwidth]{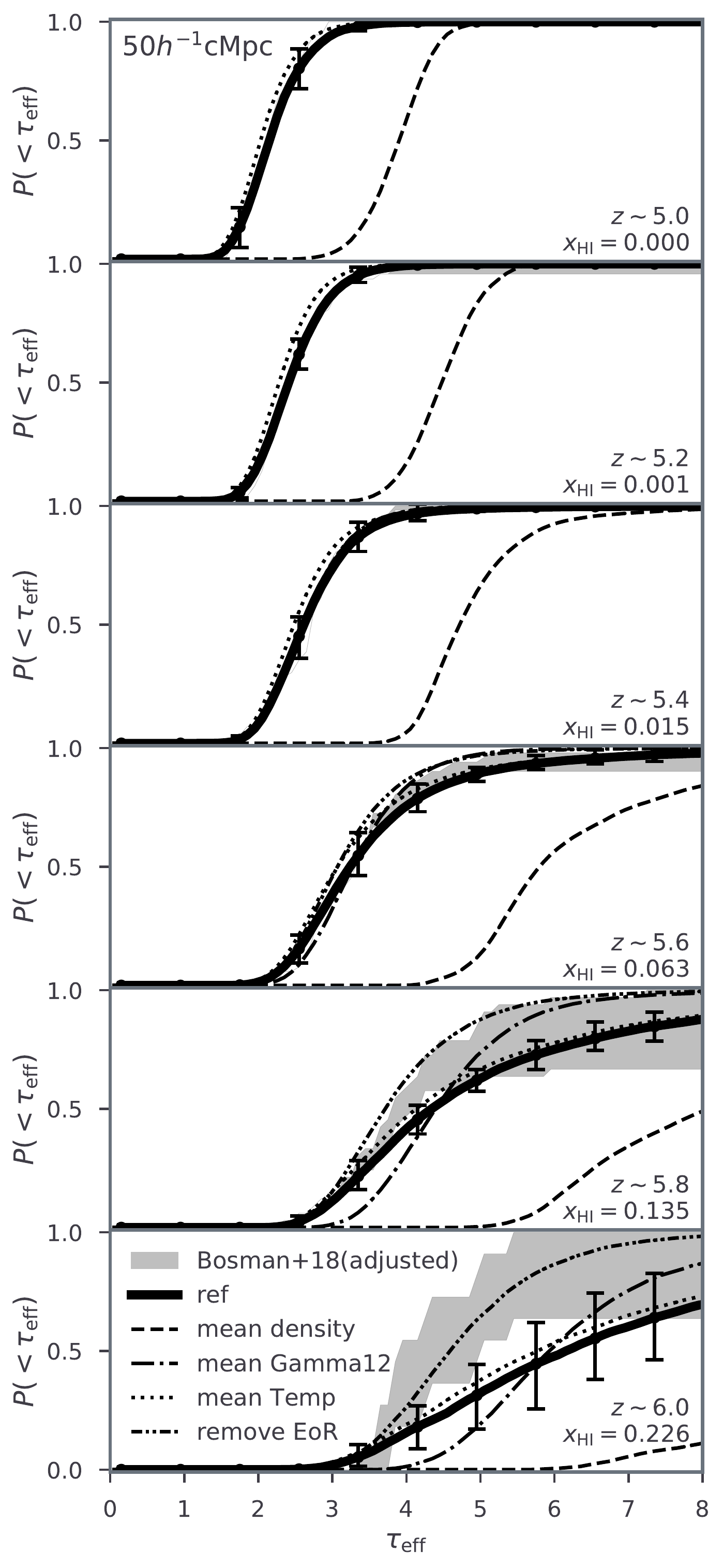}
	\end{minipage}
	\begin{minipage}{0.49\textwidth}
		\includegraphics[width=\textwidth]{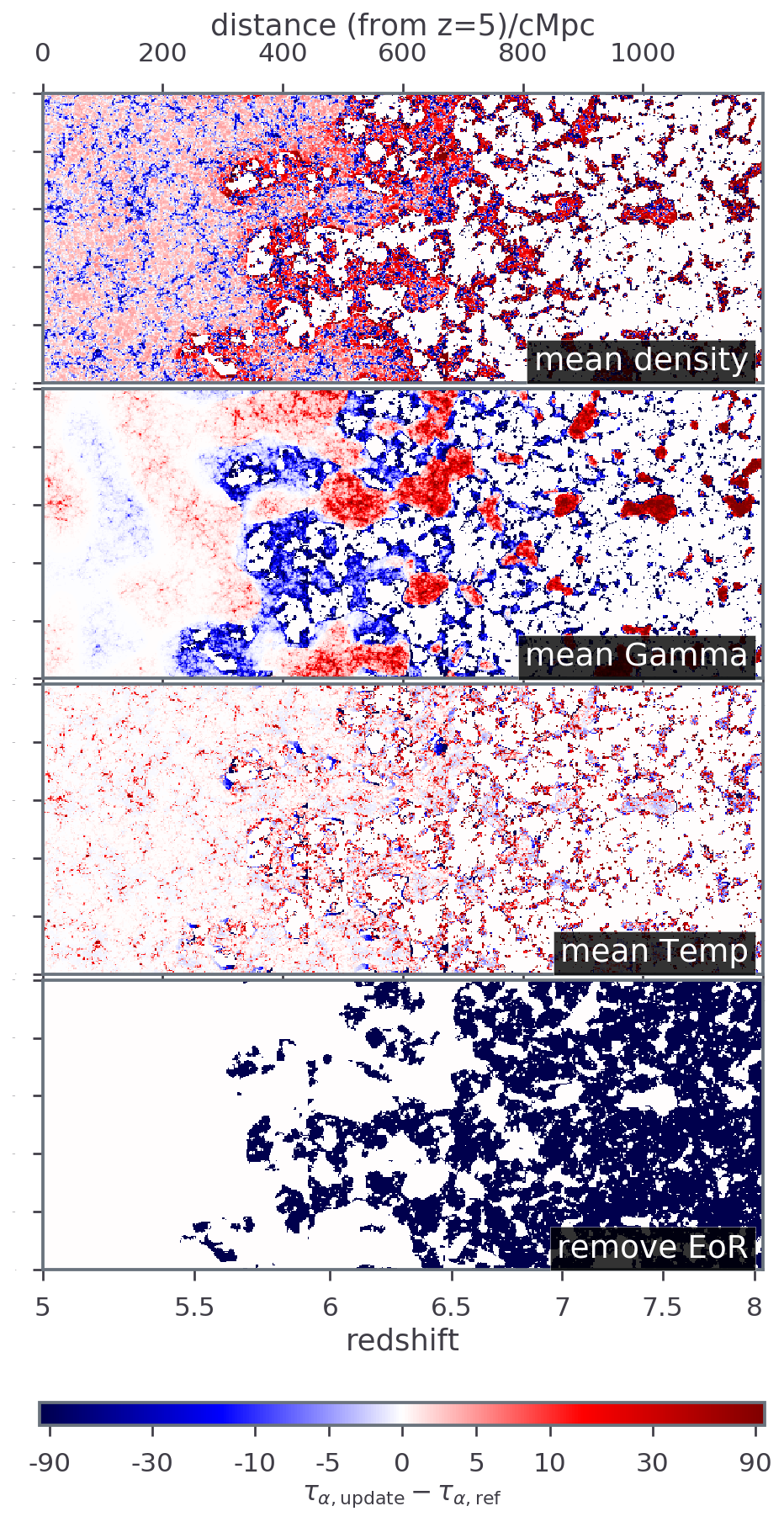}
	\end{minipage}
	\caption{\label{fig:varying_CDF}
		{\it Left  panels:} Mean CDFs of $\tau_{\rm eff}$ averaged over $50h^{-1}$cMpc at $z{=}5$--$6$ from our maximum likelihood parameter combination ({\it ref}) are shown with thick black lines, with cosmic variance uncertainties ([$14, 86$] percentiles) indicated for a subset of the bins.
		We also show CDFs resulting from removing the spatial fluctuations in various component fields (see text for details). The gray shaded regions span the observational estimates (see Sec.~\ref{sec:observation}).
		{\it Right panels:} lightcone slices illustrating the change in the {\lya} optical depth relative to the reference model for each simulation that removes fluctuations in the indicated field. Note that the colorbar is linear between $0$ and ${\pm}10$ and logarithmic from ${\pm}10$ to ${\pm}100$.  These slices illustrate that large-scale fluctuations in this model are predominantly driven by a patchy EoR and a patchy UVB.}
\end{figure*}

The panels in the figure correspond to the matter overdensity, ionizing background, temperature, neutral fraction, optical depth and \lya\ transmission, from top to bottom. We note that this ML model resulted in a rescaling factor very close to unity ($f_{\rm rescale}=0.86$--$0.96$; Section \ref{sec:parameters}). Thus, the theoretical spectra required only minor calibration using this hyperparameter, in order to be consistent with the observed data. We also list the mean values over the redshift interval $5.52{\le} z {\le} 5.88$ in the bottom panel. We see that on average {\it {\lya} Dark} has a lower ionization rate as well a lower temperature while the average densities along the two slightlines are both close to the cosmic mean.

We see that the different levels of {\lya} transmission between these two sightlines are mostly driven by the neutral hydrogen and UVB fluctuations. In the region corresponding to the long GP trough, {\it {\lya} Dark} pierces through a few remaining cosmic {$\hone$} patches (with $\Gamma_{\rm ion} \sim 0$).  Although these {$\hone$} patches combined span less than half of the GP through length, we see from the $\Gamma_{\rm ion}$ panel that the {\it ionized IGM between the {$\hone$} patches is exposed to a below-average UVB} (see also \citealt{Keating2020MNRAS.491.1736K,Nasir2020MNRAS.494.3080N}).  As discussed above, both fields correlate with the underlying density on large scales: the regions last to ionize and those with a small UVB both correspond to large-scale underdensities with comparably few star-forming galaxies. This is quantitatively evident when we compute the volume averaged density around the sightlines, showing that {\it {\lya} Dark} lies in a large-scale underdensity, $1+\overline{\delta}_{\rm 3D} =0.95$ when average over a volume of $50\times50\times110$ ($h^{-1}$ Mpc)$^3$. This picture is also consistent with follow-up observations of J0148+0600 that show a dearth of galaxies at the location of its GP trough \citep{Becker2018ApJ...863...92B,Kashino2020ApJ...888....6K}.

\subsubsection{Fields that determine {\lya} fluctuations}\label{subsec:varying fields}

Using our maximum likelihood (reference) parameter set, we now further quantify which fields are most relevant for large-scale opacity fluctuations.  In the left panels of Fig. \ref{fig:varying_CDF} we show the cumulative probability distributions (CDFs) of the effective optical depth defined as $\tau_{\rm eff}\equiv -\ln\left[\langle \exp(-\tau_{\alpha}) \rangle_{\rm L}\right]$, where we chose $L=50 h^{-1}$cMpc in order to facilitate comparison against published observations (shown as gray shaded regions and further discussed in Section \ref{sec:observation}). Thick, solid curves correspond to the mean CDFs from our ML model, for redshift bins spanning $z=$5--6.\footnote{The ML model seems to be in (very) mild tension with the data at $z=6$.  We note however that preliminary, updated data from the XQR30 large VLT program (PI: V.~D'Odorico) are in better agreement with this reference model at $z=6$.}.  The other curves in these panels were constructed by removing fluctuations in a single component field, and then recomputing the resulting optical depths (though keeping the same normalization factor, $f_{\rm rescale}$).  Specifically, we replace the density ({\it mean density}), photo-ionization rate ({\it mean Gamma}) and temperature ({\it mean Temp}) with their ionized volume averaged values at each redshift.  For {\it remove EoR}, we remove all cosmic {$\hone$} patches by assigning to them the mean values of the photo-ionization rate and temperature in the ionized regions, and then recomputing $\tau_\alpha$ in each cell.

In the right panels, we also show the lightcone visualizations of the corresponding change in the optical depth with respect to the reference ML model (see the transmission in Fig.~\ref{fig:reference}).  As expected, ignoring density fluctuations results in a large shift in the CDFs, but does not dramatically impact their shapes -- there is only a minor decrease in the abundance of high $\tau_{\rm eff}$ regions.  Thus density fluctuations alone cannot efficiently generate long GP troughs \citep{Becker2015MNRAS.447.3402B}).  Ignoring the temperature fluctuations has a minor impact, mostly on small scales.

The most striking difference comes from the patchy EoR and UVB fields.  We see that the fluctuations in both fields have a relatively large impact on the $\tau_{\rm eff}$ CDFs during the final stages of the EoR ($z \gsim 5.5$), affecting the opacity on large scales.  In agreement with the discussion in the previous section (see also e.g. \citealt{Keating2020MNRAS.491.1736K,Nasir2020MNRAS.494.3080N}), their impact is additive -- they both widen the CDFs, extending the high-value tails of $\tau_{\rm eff}$.  Post EoR as the mean free path increases, the fluctuations in the UVB become negligible when computing the $\tau_{\rm eff}$ CDFs\footnote{We caution that our model does not include AGN.  AGNs are expected to dominate the UVB at lower redshifts, $z\lsim3$--$4$ (e.g.~\citealt{Madau2015,Mitra2015,Qin2017a,Kulkarni2019MNRAS.485L..24K}); however, some works have evoked rare, bright AGN to explain the observed large scale $\tau_{\rm eff}$ fluctuations at $z\sim5$--6 (e.g.~\citealt{Chardin2017MNRAS.465.3429C,Meiksin2020MNRAS.491.4884M}).  Regardless, our models are able to fit the data at $z>5$ without AGN (see also, e.g.~\citealt{Nasir2020MNRAS.494.3080N,Keating2020MNRAS.491.1736K}).}.

\section{Observational data}\label{sec:observation}

We constrain our model using the most up to date, public {\lya} forest sample\footnote{\url{www.sarahbosman.co.uk/research}} from \citet{Bosman2018MNRAS.479.1055B}.
Specifically, we use the ``SILVER'' subsample of $51$ QSO that have high quality spectra with signal-to-noise ratio (S/N) ${\ge}5.3$ in the continuum. The spectra are binned over $50h^{-1}$cMpc, resulting in $217$ flux measurements over the redshift interval $4.9{\le} z {\le} 6.1$ (i.e.~$\langle F\rangle_{50h^{-1}{\rm cMpc}}$ with uncertainties of $\delta_{\langle F\rangle_{50h^{-1}{\rm cMpc}}}$). We then have the following number of samples, $N_{\rm sample}$= $18$, $47$, $57$, $51$, $33$ and $11$ at $z = 5.0$, $5.2$, $5.4$, $5.6$, $5.8$ and $6.0$, respectively.
The effective optical depth is then calculated as $\tau_{\rm eff} {=} {-}\ln\langle F\rangle_{50h^{-1}{\rm cMpc}}$ and, for non-detections, $\tau_{\rm eff} {\ge} {-}\ln\left(2\delta_{\langle F\rangle_{50h^{-1}{\rm cMpc}}}\right)$ as the lower limit\footnote{Note that a few objects in \citet{Bosman2018MNRAS.479.1055B} have different lower limits from $\tau_{\rm eff} {\ge} {-}\ln\left(2\delta_{\langle F\rangle_{50h^{-1}{\rm cMpc}}}\right)$ as they adopt the peak limit definition proposed by \citet{Becker2015MNRAS.447.3402B}.}.

We then rescale these effective optical depths to account for the improved continuum reconstruction from \citet{Bosman2020arXiv200610744B}.
Specifically, we use the PCA-nominal method from \citet{Bosman2020arXiv200610744B}, scaling the transmission by the ratio of the two continuum estimates,  $\tau_{\rm eff}{=}{-}\ln\left(\langle F\rangle_{50h^{-1}{\rm cMpc}}\mathscr{R}\right)$, and the lower limits to $\tau_{\rm eff} {\ge} {-}\ln\left(2\delta_{\langle F\rangle_{50h^{-1}{\rm cMpc}}}\mathscr{R}\right)$ for non-detections. Here, $\mathscr{R}\equiv\langle F \rangle_{\rm obs}^{\rm Bosman+20}/ \langle F \rangle_{\rm obs}^{\rm Bosman+18}$ represents the mean flux ratio between \citet{Bosman2018MNRAS.479.1055B} and \citet{Bosman2020arXiv200610744B}, which is $0.93$, $0.87$, $0.91$, $0.87$, $1.03$ and $1.60$ at $z=5.0$, $5.2$, $5.4$, $5.6$, $5.8$ and $6.0$, respectively. We note that this only has a noticeable change to the CDF of $\tau_{\rm eff}$ at $z=6$ where the peak is at a slightly lower value.

The resulting observed CDFs of $\tau_{\rm eff}$ are shown as the gray shaded regions in Fig. \ref{fig:varying_CDF}. When performing EoR inference from these observational estimates, we consider flat likelihoods between the two extremes (i.e.~the gray shaded region; see more in Section \ref{subsec:step}).

\section{Summary of model parameters}\label{sec:parameters}

Before proceeding to our MCMC results, we summarize the free parameters of our model and the associated prior ranges.  For computational convenience in this proof-of-concept work we restrict our parameter exploration to the most relevant astrophysical parameters (i.e.~responsible for the largest variation in the forward-modelled data); for a complete list of parameters see Section 2 and \citet{Park2019MNRAS.484..933P}.  We also do not co-vary cosmological parameters, keeping the same underlying density field.  In future work, we will relax this assumption and quantify joint constraints on astrophysical and cosmological parameters from the {\lya} forest.

Our model consists of six astrophysical parameters characterizing the UV emission of galaxies:
\begin{enumerate}
	\item $f_{*,10}$, the fraction of galactic baryons inside stars, defined for galaxies residing in halos with $M_{\rm vir}=10^{10}{\msol}$. We vary $f_{*,10}$ between $10^{-3}$ and 1 with a flat prior in log space;
	\item $\alpha_*$, the power-law index of the stellar fraction to halo mass relation. We vary it between $-0.5$ and $1$ with a flat prior, noting that high-redshift observations of galaxy UV LFs suggest $\alpha_*{\sim}0.5$.
	\item $f_{\rm esc,10}$, the UV ionizing escape fraction, defined for galaxies residing in halos with $M_{\rm vir}=10^{10}{\msol}$. Similarly to $f_{*,10}$, we vary $f_{\rm esc,10}$ between $10^{-3}$ and 1 with a flat prior in log space;
	\item $\alpha_{\rm esc}$, the power-law index of the ionizing escape fraction to halo mass relation. Although poorly known, some works suggest increasing or constant escape fractions towards lower mass galaxies (\citealt{Ferrara2013MNRAS.431.2826F,Kimm2014ApJ...788..121K,Paardekooper2015MNRAS.451.2544P,Xu2016ApJ...833...84X}).  We thus have a wider prior on the negative range, varying $\alpha_{\rm esc}$ between $-1$ and $0.5$ (with a flat prior);
	\item $M_{\rm turn}$, the turnover halo mass below which the number density of halos hosting star-forming galaxies become exponentially suppressed. We  vary $M_{\rm turn}$ between $10^8$ and $10^{10}{\msol}$ with a flat prior in log space. The former corresponds to the atomic cooling threshold while the latter corresponds to current {\it HST} sensitivity limits beyond which we see a high occupancy fraction of star forming galaxies; and
	\item $t_*$, the star-formation timescale as a fraction of the Hubble time. We vary $t_*$ between 0 and 1 with a flat prior.
\end{enumerate}

We also introduce two hyper-parameters to describe the optical depth normalization factor, $f_{\rm rescale}$ from equation (\ref{eq:tau_lya}). Hyper (or nuisance) parameters can characterize systematics and errors, and are marginalized over to obtain the final constraints on the desired parameters.  As mentioned in the associated discussion, the $f_{\rm rescale}$ normalization factor can account for errors in the continuum subtraction as well as modelling errors.  However, it comes with the cost of losing the intrinsic predictive power of the model.  In order to preserve some of the predictive power of our model, we assign these hyper-parameters a prior centered around $f_{\rm rescale} = 1$ (with unity corresponding to no calibration).  The chosen widths are fairly arbitrary, with the fiducial \citet{Park2019MNRAS.484..933P} model included within $1\sigma$; however, below we also explore an extreme case that essentially allows infinitely wide priors.  Specifically, the two hyper-parameters of our fiducial MCMC are:
\begin{enumerate}
	\item[(vii)] $f_{\rm rescale}(z{=}5.7)$, the rescaling factor at $z=5.7$, sampled in log space between $10^{-0.5}$ and $10^{0.5}$ with a Gaussian prior of a zero mean (i.e. $\langle\log_{10}f_{\rm rescale}(z=5.7)\rangle=0$) and a width of $\sigma \log_{10}f_{\rm rescale}(z{=}5.7)=1.5$;
	\item[(viii)] ${\rm d}f_{\rm rescale}/{\rm d}z$, the slope of the rescaling factor as a function of redshift, sampled between -1.5 and 1.5 with a Gaussian prior of a zero mean (i.e. $\langle{\rm d}f_{\rm rescale}/{\rm d}z\rangle=0$) and a width of $\sigma {\rm d}f_{\rm rescale}/{\rm d}z=2.0$.
\end{enumerate}

\section{Inferring galaxy and IGM properties from the {\lya} forest}
\label{sec:inference}

In this section, we use the {\cmmc} sampler \citep{Greig2015MNRAS.449.4246G,Greig2017MNRAS.472.2651G,Greig2018MNRAS.477.3217G} to perform three MCMC runs, quantifying EoR and galaxy constraints with and without the \lya\ forest data.  For computational efficiency, our MCMCs are done on smaller volumes ($250$ cMpc on a side) with the same cell size as the example shown in Section \ref{sec:models} (i.e.~cell lengths of $250/128 = 1.95$ cMpc).  Our MCMC runs typically require $\sim10^5$ samples to reach convergence, which take $\sim15000$ CPU hours.

\subsection{Computing the likelihood}\label{subsec:step}

For a given combination of galaxy and hyperparameters, $\theta$,
we compute the associated likelihood from Ly$\alpha$ forest fluctuations, $\mathscr{L}_{\alpha}(\theta)$, according to the following:

\begin{enumerate}
	\item We forward model the corresponding 3D lightcone of the {\lya} optical depth as described in Section \ref{subsec:forest}.
	
	\item At a given redshift $z$, we compute the PDF\footnote{As noted by \citet{Choudhury2020arXiv200308958C}, we use PDFs instead of the more commonly presented CDFs when computing the likelihood, as CDFs have stronger covariances between bins.}	of the effective optical depths from a sample of $N_{\rm sample}$ maximally-separated (to minimize spatial coherence) sightlines, with $N_{\rm sample}$ corresponding to the sample size of the observational data at $z$ as discussed in Sec. \ref{sec:observation}.

 	\item We repeat step (ii) $N_{\rm realization}{=}150$ times, ensuring each sightline in the lightcone is not selected more than once.  From these realizations we compute the mean PDF ($\phi_{\rm model}$), and the cosmic variance error matrix ($\Sigma_{\rm CV}$; see more in Appendix \ref{app:sec:ECM}).
 	
 	\item We calculate the total error covariance matrix\footnote{We do not include flux uncertainties in the total covariance matrix, as we verify they are far subdominant compared to the FGPA and cosmic variance errors for this observational sample. In future work, we will extend our framework and forward model each individual sightline, including the corresponding flux errors.} as $\Sigma=\Sigma_{\rm CV}+ \Sigma_{\rm GP}$, where $\Sigma_{\rm GP}$ corresponds to the error from the FGPA (see Appendix \ref{app:sec:FGPA}). 

	\item We calculate the difference\footnote{As mentioned in Section \ref{sec:observation}, we consider flat likelihoods between the two extreme cases shown in Fig. \ref{fig:varying_CDF}. For instance, assuming the observed number density in a particular $\tau_{\rm eff}$ bin has the limits $\phi_{\rm obs,upper}$ and $\phi_{\rm obs,lower}$, we take $X{=}0$ when $\phi_{\rm obs,lower}{\le}\phi_{\rm model}{\le}\phi_{\rm obs,upper}$, and $X{=}\min\left(|\phi_{\rm model}{-}\phi_{\rm obs,upper}|, |\phi_{\rm model}{-}\phi_{\rm obs,lower}|\right)$ otherwise.} between the modelled mean PDF and the observed PDF, $X$.
	
	\item From (iv) and (v) we compute a $\chi^2$ likelihood for this redshift, according to $\ln \mathscr{L}_{z}(\theta) = -0.5 X^T \Sigma^{-1}X$.
	
	\item We repeat steps (ii) to (vi) for all redshift bins used in our analysis\footnote{We expect our models to become less accurate at lower redshifts since, (i) we do not account for the contribution of AGN to the UV background; and (ii) we do not capture the spatial clustering of biased absorbers such as DLAs.  As discussed above, we expect these approximations to be reasonable at $z>5$; however for this proof-of-concept work, we conservatively restrict ourselves to the highest redshift bins that are the most sensitive to the EoR. In future works, we will explore extending this range, including an additional population of bright AGN.}  ($z=5.4, 5.6, 5.8$ and 6.0), summing up the log likelihoods: $\ln \mathscr{L}_{\alpha}(\theta) = \sum_{z} \ln \mathscr{L}_{z}(\theta)$.

\end{enumerate}

Finally, we obtain the total likelihood with: $\mathscr{L}(\theta) = \mathscr{L}_{\alpha} \times \mathscr{L}_{\rm LFs} \times \mathscr{L}_{\rm DF} \times \mathscr{L}_{\tau_e}$, where the final three terms correspond to current, robust EoR constraints from: (i) the galaxy UV LFs at $z{=}6{-}10$ from \citet{Bouwens2015a,Bouwens2016} and \citet{Oesch2018ApJ...855..105O}; (ii) the upper limit on the neutral hydrogen fraction at $z\sim5.9$, $x_{\hone}<0.06{+}0.05(1\sigma)$, measured using the dark fraction of quasar spectra \citep{McGreer2015MNRAS.447..499M}; and
(iii) the Thomson scattering optical depth of CMB photons reported by \citet{Planck2018arXiv180706209P}, $\tau_e=0.0561{\pm}0.0071(1\sigma)$;
respectively.
For more details about these observations and the functional forms of the corresponding likelihoods, see \citet{Park2019MNRAS.484..933P}.

\subsection{Posteriors}

Fig. ~\ref{fig:MCMC} presents the MCMC results including: the marginalized  posterior distributions of the model parameters (the corner plot in the lower left), the EoR histories (panels a), $\tau_{\rm eff}$ distributions at $z{=}5.4-6.0$ (i.e.~CDF in panels b), galaxy UV LFs at $z{=}6-15$ (panels c), the UVB evolution (panel d) as well as the PDF of the CMB Thomson scattering optical depth, $\tau_e$ (panel e).

We perform the following three MCMC runs:
\begin{itemize}
\item \textbf{\textit{no\_forest}} -- does not use the $\tau_{\rm eff}$ PDFs, with the likelihood corresponding to $\mathscr{L}(\theta) = \mathscr{L}_{\rm LFs} \times \mathscr{L}_{\rm DF} \times \mathscr{L}_{\tau_e}$. This run roughly corresponds to our current state of knowledge, before accounting for \lya\ opacity fluctuations.  Here, we vary the astrophysical parameters (i) -- (vi) from Sec. \ref{sec:parameters}.

\item \textbf{\textit{forest}} -- additionally includes the observed $\tau_{\rm eff}$ PDFs discussed in Sec. \ref{sec:observation}; thus $\mathscr{L}(\theta) = \mathscr{L}_{\alpha} \times \mathscr{L}_{\rm LFs} \times \mathscr{L}_{\rm DF} \times \mathscr{L}_{\tau_e}$.  Here, we vary the astrophysical and nuisance parameters, (i) -- (viii) from Sec. \ref{sec:parameters}.  {\it This corresponds to our fiducial MCMC run.}

\item \textbf{\textit{forest\_fluc}} -- unlike \textit{forest}, does not sample the $f_{\rm rescale}$ hyper-parameters from their priors.  Instead we normalize each PDF by insuring the mean flux, $\langle \exp\left({-\tau_{\alpha}}\right)\rangle$, matches the observed mean flux. This roughly corresponds to assuming infinitely wide priors for $f_{\rm rescale}$. Therefore, this run only varies the astrophysical parameters (i) -- (vi) from Sec. \ref{sec:parameters}.
\end{itemize}
Below we discuss the posterior of each of these in turn.

\subsubsection{\textit{no\_forest}}

The {\it no\_forest} posterior is shown with the orange, shaded regions.  As noted by \citet{Park2019MNRAS.484..933P}, current EoR observations already place constraints on some of our model parameters, even without making use of the $\tau_{\rm eff}$ PDF data.  In particular, galaxy LFs constrain the stellar to halo mass relation, parametrized here through $f_{*,10}/t_*$ and $\alpha_\ast$, to within a factor of few (see also, e.g.~\citealt{Tacchella2018ApJ...868...92T,Behroozi2019MNRAS.488.3143B,Yung2019MNRAS.490.2855Y,Qin2020arXiv200616828Q}).  This is also evident from the tight recovery of the UV LFs at the bright end ($M_{\rm UV} \lsim -15$).  However, we do not detect a faint-end turnover in the LFs, resulting from inefficient star formation in galaxies hosted by halos with masses less than $M_{\rm turn}$; current UV LFs only provide upper limits on this parameter (${\lesssim}5\times10^9\msol$; e.g.~\citealt{Bouwens2015b}).
The ionizing escape fraction is only weakly constrained by the dark fraction and CMB limits on EoR timing: $f_{\rm esc,10}{\sim}4$--26 per cent), while its scaling with the mass of the host halo remains uninformed (see also e.g. \citealt{Haardt2012,Kuhlen2012,Robertson2015ApJ...802L..19R,Price2016}).

\subsubsection{\textit{forest}}

\begin{figure*}
	\begin{minipage}{\textwidth}
		\begin{center}
			\vspace*{-3mm}
			\includegraphics[width=1.\textwidth]{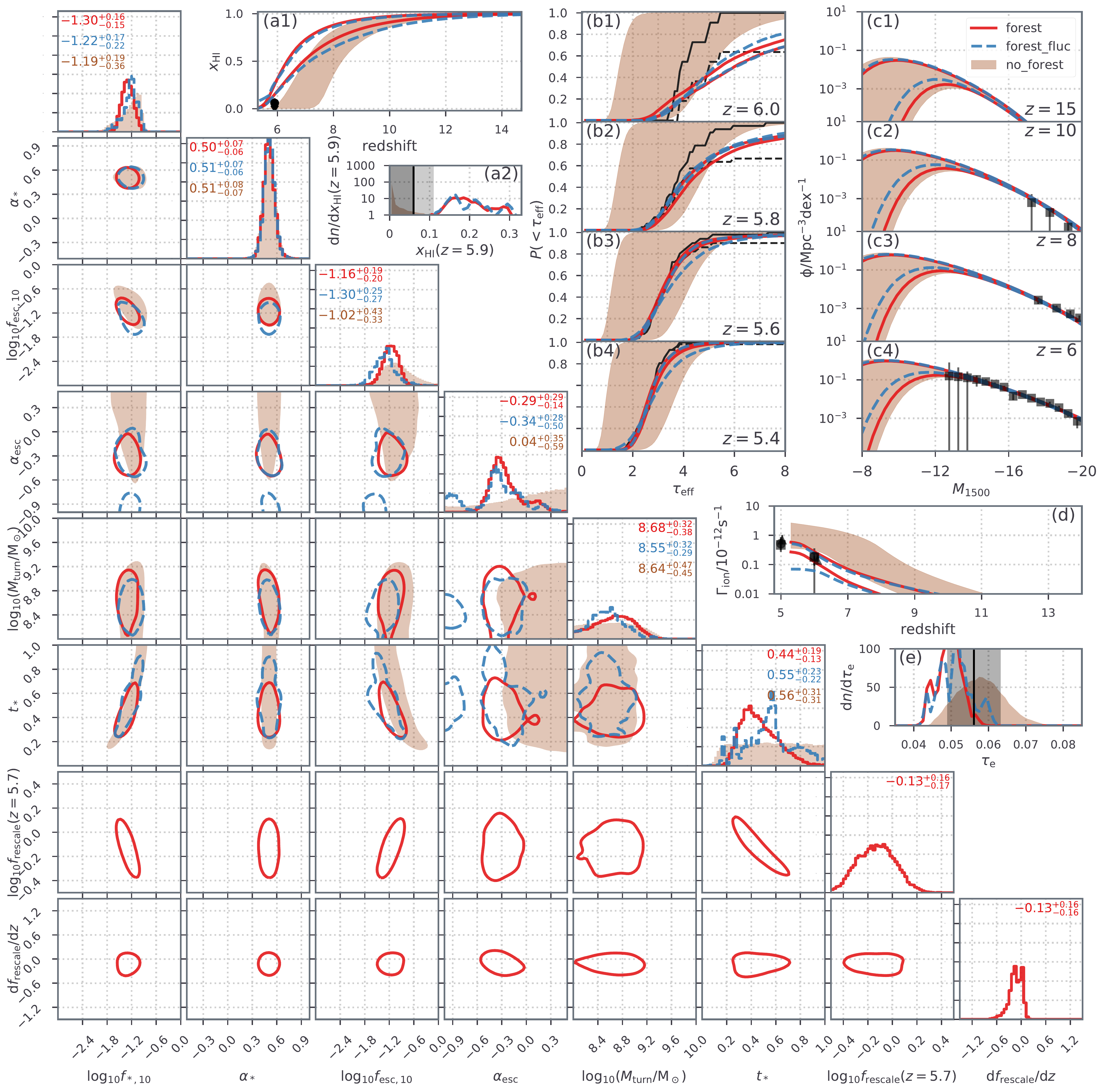}
		\end{center}
	\end{minipage}
	\caption{\label{fig:MCMC}Marginalized posterior distributions of the 
		model parameters with ({\it forest}; red solid lines \& {\it forest\_fluc}; blue dashed lines) and without ({\it no\_forest}; brown shaded areas) the forest data \citep{Bosman2018MNRAS.479.1055B}.
		All three results use the observed galaxy LFs at $z=6-10$ 
		\citep{Bouwens2015a,Bouwens2017ApJ...843..129B,Oesch2018ApJ...855..105O}, upper 
		limits on the neutral fraction at $z\sim5.9$ from quasar spectra 
		\citep{McGreer2015MNRAS.447..499M}, and Thomson scattering optical depth of the CMB 
		\citep{Planck2018arXiv180706209P} as input constraints. 
		The 2D distributions correspond to the 68th percentiles.
		The medians with [$14, 86$] percentiles for each parameter are presented in the upper corner of the 1D PDFs (from top to bottom: {\it forest}, {\it forest\_fluc} and {\it no\_forest}, respectively).
		Note that all 1D PDFs have been normalized to have area (or integral) under the histogram equal to 1.
		The upper right sub-panels present the median and [14, 86] 
		percentiles of the neutral hydrogen ($x_{\hone}$; panel a1); the CDFs of {\lya} effective optical depths at $z=5.4-6.0$ (panels b1-b4); the UV LFs at $z=6-15$ (panels c1-c4); the evolution of the photoionization rate ($\Gamma_{\rm ion}$; panel d); as well as the PDFs of $\tau_e$ (panel e) and 
		$x_{\hone}$ at $z=5.9$ (panel a2) for the models presented in the posterior 
		distributions. Observations including UV ionizing background measured by \citet{Bolton2007MNRAS.382..325B,Calverley2011MNRAS.412.2543C,Wyithe2011MNRAS.412.1926W} are indicated in black.}
\end{figure*}

Additionally including the $\tau_{\rm eff}$ data has a dramatic impact on the posterior (shown with the red curves in Fig.~\ref{fig:MCMC}).  Most importantly, we note from panel (a1) that {\it the forest data requires late reionization}.  The final overlap stages of the EoR, corresponding to when ${>}95$\% of the volume was ionized, occur at $z{<}5.6$.  
Our Bayesian framework provides statistical proof of previous suggestions that the EoR might have completed at $z{<}6$ \citep{Lidz2007ApJ...670...39L,Mesinger2010MNRAS.407.1328M,Kulkarni2019MNRAS.485L..24K,Keating2020MNRAS.491.1736K,Nasir2020MNRAS.494.3080N}.
Our constraints on overlap at $z{<}5.6$ are also perfectly consistent with the recent, similar analysis by \citet{Choudhury2020arXiv200308958C}, using their indirect parametrization of ionizing sources and IGM recombinations.

Moreover, we find that the $\tau_{\rm eff}$ PDFs and the dark fraction upper limits are in mild tension ($\sim 1.5\sigma$; see panel a2). The dark fraction tends to prefer earlier reionization, while the $\tau_{\rm eff}$ PDFs require later reionization. Combining the two forces the marginalized posterior of EoR histories to have a very narrow tail bellow $z<6$ (see panel a1).  We will revisit this mild tension in future work, using updated estimates of both data sets from the XQR30 large observational program (PI: V.~D'Odorico).

As it helps nail down the timing of the EoR, the forest data dramatically improves constraints on the ionizing escape fraction. To be consistent with all of the data, reionization would need to end late (preferred by the broad $\tau_{\rm eff}$ PDFs) but not {\it too} late or with an extended tail towards low redshifts (preferred by the modest values of $\tau_e$ and the dark fraction). These limit the ionizing contribution of both early-forming, faint galaxies and their late-appearing, bright counterparts, resulting in $\alpha_{\rm esc}$ that peaks around $-0.29^{+0.29}_{-0.14}$, a characteristic ionizing escape fraction of $f_{\rm esc, 10}= 6.9^{+3.8}_{-2.6}$\%, as well as a weakly constrained ($\sim 1 \sigma$) lower limit on $M_{\rm turn}$ (i.e.~${\gtrsim}2{\times}10^8\msol$).

As we mentioned previously, in this initial study we do not enforce UV photon conservation, in favor of simulation speed (setting \textsc{photon\_cons=false} in {\tocm}v3).  This choice roughly leads to a $+0.2$ bias in the inferred $\alpha_{\rm esc}$ parameter (Park et al. in prep). Taking this into account, we predict that the true marginalized posterior of $\alpha_{\rm esc}$ is consistent with zero and has a modest width of $\sigma_{\alpha_{\rm esc}}\lsim 0.3$. This suggests that {\it the forest data disfavors a strong evolution of the ionizing escape fraction with the galaxy's halo (or stellar) mass} (i.e.~large $\lvert \alpha_{\rm esc} \rvert$), consistent with recent results from hydrodynamical simulations of a handful high-$z$ galaxies in the relevant mass ranges \citep{Kimm2014ApJ...788..121K,Ma2015MNRAS.453..960M,Xu2016ApJ...833...84X,Ma2020MNRAS.498.2001M}.

Finally, it is useful to point out a few sanity checks of our model.  Firstly, we see that the recovered hyperparameters, accounting for systematic errors in continuum subtraction and theoretical modelling, are consistent with $f_{\rm rescale} = 1$.  This implies that our intrinsic models of the forest can match the forest data, {\it without significant ``tuning''}.  Secondly, we note from panel (d) that our {\it forest} posterior matches perfectly with estimates of the mean ionizing background at $z=5$ and 6 (denoted by black points with error bars; \citealt{Bolton2007MNRAS.382..325B,Calverley2011MNRAS.412.2543C,Wyithe2011MNRAS.412.1926W}).
It is worth noting that, although they are based on forest observations, {\it we do not use these estimates of the UVB in our likelihood}.

\subsubsection{\textit{forest\_fluc}}

The {\it forest\_fluc} posterior is denoted with blue curves in Fig.~\ref{fig:MCMC}.  Recall that here we calibrate each forward modelled $\tau_{\rm eff}$ PDF such that the mean flux matches the data, for every redshift bin and every parameter sample.  This corresponds to an extreme case in which we have no idea how to put priors on $f_{\rm rescale}$.  In effect, we are removing the ``DC mode'' of our forest models, mostly comparing the shapes of the PDFs over the observable range.

\begin{figure*}
	\begin{minipage}{\textwidth}
		\centering
		\includegraphics[width=\textwidth]{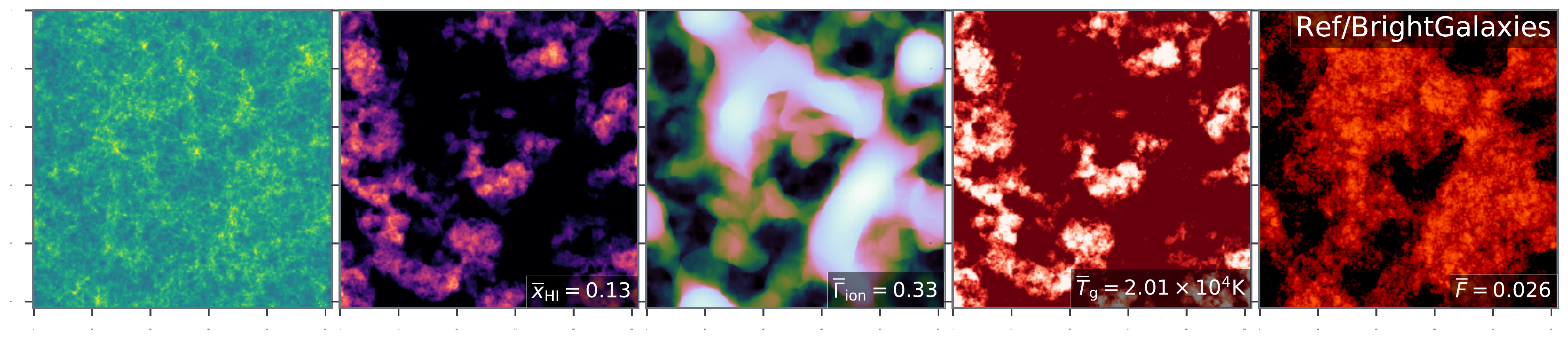} \vspace*{-8mm}\\
		\includegraphics[width=\textwidth]{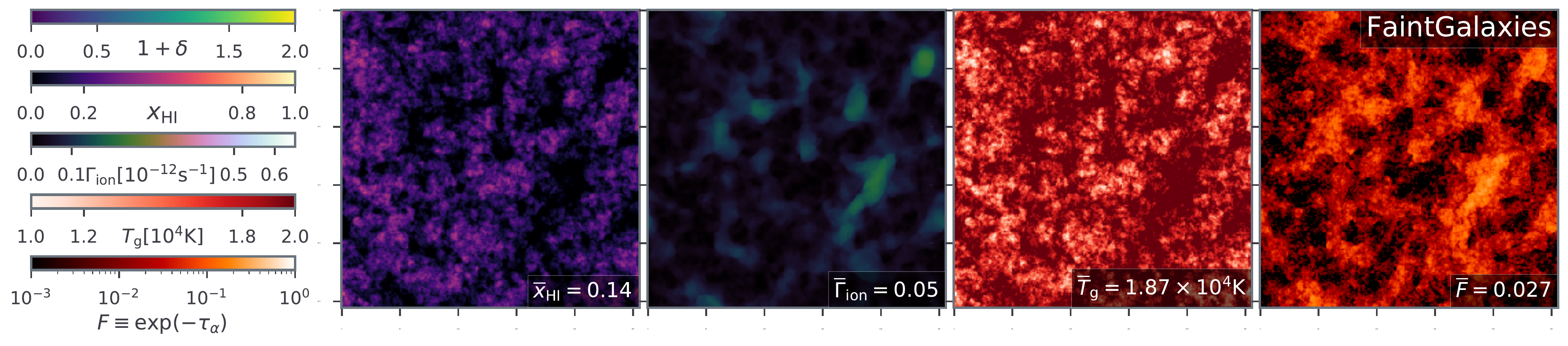}
	\end{minipage}
	\caption{\label{fig:tau_alpha_comparisons} Slices (with a side length of $50h^{-1}$cMpc and depth of 2 cMpc) of the density ($1+\delta$), neutral hydrogen fraction ($x_{\hone}$), photonionization rate ($\Gamma_{\rm ion}$), gas temperature ($T_{\rm g}$) and {\lya} transmission flux from {\it Ref/BrightGalaxies} (upper subpanels) and {\it FaintGalaxies} (lower ones); see text for more details on these models.  All slices correspond to $z=5.8$ where the two models have the same filling factor of {$\hone$}, and were rerun in larger volumes (500 cMpc) compared to the MCMC for better visualization.}
\end{figure*}

Even when calibrating to the observed mean flux, we recover the result that EoR must finish at $z<5.6$ to match the observed {\it widths} of the Ly$\alpha$ opacity distributions (panel a1). By comparing the blue curves to the red ones, we see that losing information on the mean flux expands the recovered posterior to include models with a very low ionizing background (see panel d).  These are mostly sourced by small values of $\alpha_{\rm esc}$, corresponding to enhanced ionizing efficiencies in galaxies hosted by low-mass halos.  The corresponding EoR histories are also somewhat slower, since the fractional growth of less massive halos is slower than that of more massive ones.  As a result, the marginal posterior on $\alpha_{\rm esc}$ is bi-modal, with the smaller peak at low values being driven by a slightly better agreement with $\tau_e$ from {\it Planck} due to the more extended EoR.

\subsection{Can we distinguish between different reionization morphologies?}\label{subsec:tau_alpha_comparisons}

In the previous section, we demonstrated that our models require reionization to be incomplete at $z<6$ in order to match the observed $\tau_{\rm eff}$ PDFs, while in Sec.~\ref{subsec:varying fields} we saw that both the patchy UVB and patchy EoR are needed to reproduce the longest GP troughs (see also \citealt{Keating2020MNRAS.491.1736K}). In this section, we check if the forest opacity fluctuations can distinguish between different reionization morphologies, {\it at a fixed stage of the EoR}.  Constraining the bias of the dominant galaxy population through the EoR morphology would be immensely powerful, allowing us to distinguish between different galaxy models that result in similar reionization histories (e.g.~\citealt{McQuinn2007MNRAS.377.1043M,Dixon2016MNRAS.456.3011D,Mesinger2016,Ahn2020arXiv201103582A}). 
Indeed, the ability to measure the morphology of the EoR and epoch of heating is one of the main reasons the cosmic 21-cm signal will allow us to place $\sim$ percent level constraints on the properties of the unseen first galaxies (e.g.~see the recent review in \citealt{Mesinger2019cosm.book.....M}).

\begin{figure}
	\begin{minipage}{\columnwidth}
		\includegraphics[width=\textwidth]{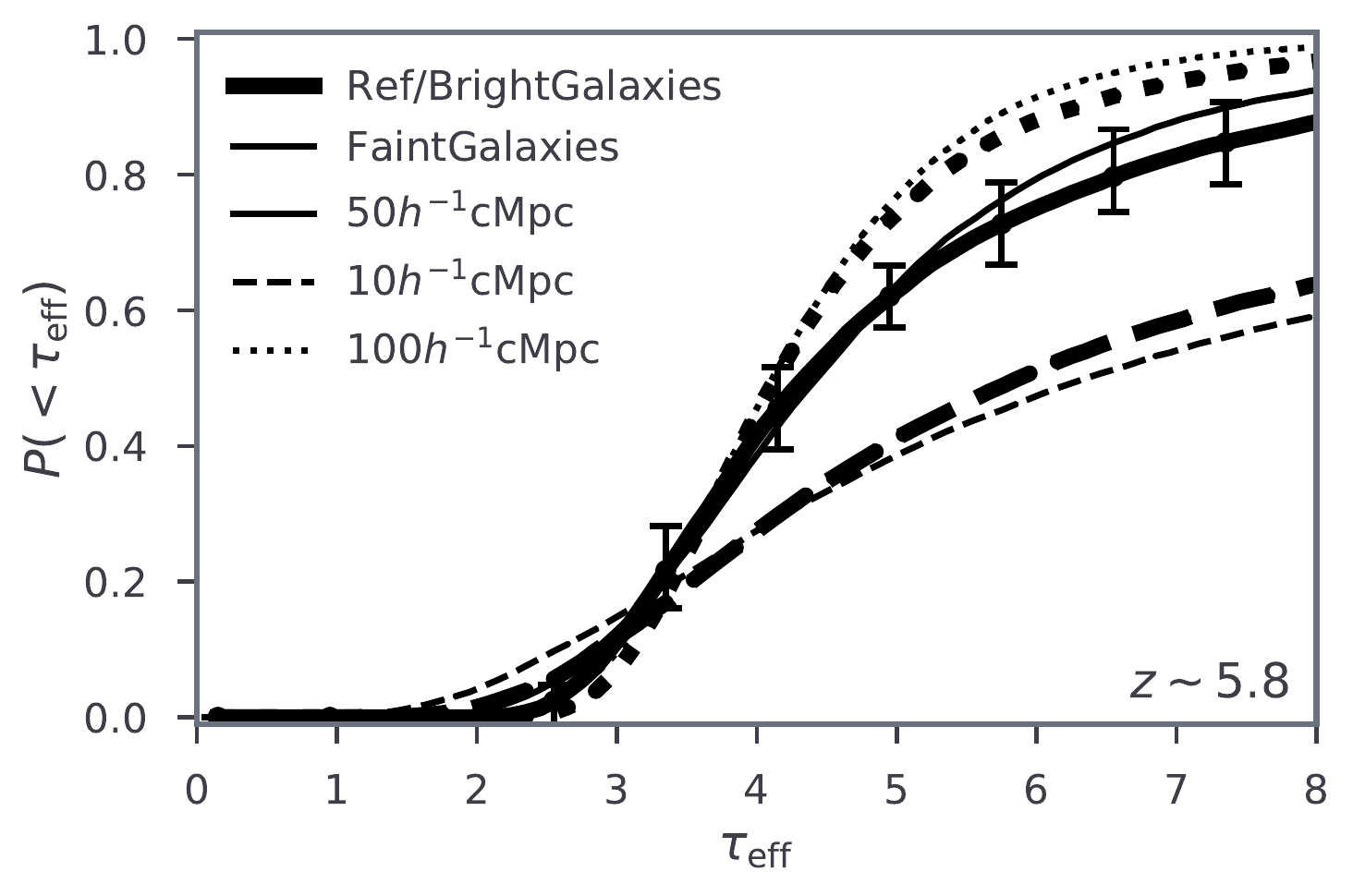}
	\end{minipage}
	\caption{\label{fig:tau_alpha_comparisons_CDF} CDFs of $\tau_{\rm eff}$ and $\ln\tau_{\rm eff}$ at $z=5.8$ averaged over 50, 10 and $100h^{-1}$cMpc from {\it Ref/BrightGalaxies} (thick curves) and {\it FaintGalaxies} (thin curves). Uncertainties ([14, 86] percentiles) for the $50h^{-1}$cMpc result in {\it Ref/BrightGalaxies} are indicated in a subset of the bins.}
\end{figure}

In Fig.~\ref{fig:tau_alpha_comparisons}, we show slices through the fields of two models at $z=5.8$, chosen to have the same neutral fraction $(\bar{x}_{\hone} {\sim}0.13)$ but very different EoR morphologies.  The top panels correspond to our reference (ML) model, discussed already in Sec. \ref{subsec:forest}. This model has $\alpha_{\rm esc}{=}-0.298$ and $M_{\rm turn}{=}7.2\times10^8\msol$, resulting in relatively massive galaxies driving the EoR; thus we label it {\it Ref/BrightGalaxies}.  Below it we show a {\it FaintGalaxies} model with the following astrophysical parameters ($f_{*,10}$, $\alpha_*$, $f_{\rm esc,10}$, $\alpha_{\rm esc}$, $M_{\rm turn}$, $t_*$) = ($0.084$,$0.483$, $0.019$, $-0.834$, $2.16\times10^{8}{\msol}$, $0.678$). Note that we also include $f_{\rm rescale}{=}0.12$ in the {\it FaintGalaxies} model to rescale the {\lya} transmission at $z{=}5.8$ (see equation \ref{eq:tau_lya}; compared to $f_{\rm rescale}{=}0.92$ in {\it BrightGalaxies}). Although both models have the same H{\small{II}} filling factor at $z=5.8$, the EoR morphologies are noticeably different, as expected from the different biases of their corresponding dominant galaxy population.  Specifically, {\it FaintGalaxies} is characterized with more numerous, smaller H{\small{II}} regions driven by its more abundant, yet fainter galaxies. As a result, the UVB is weaker, as the contribution of very distant sources is limited by the small sizes of the H{\small{II}} regions. 

To quantify if the different EoR morphologies result in different $\tau_{\rm eff}$ distributions, in Fig.~\ref{fig:tau_alpha_comparisons_CDF} we compute the CDFs averaging over $10$, $50$ (the default value used in the MCMC), and $100 h^{-1}$Mpc\footnote{We note that the {\it FaintGalaxies} model has a different EoR evolution and is disfavored by the {\it total} likelihood in {\it forest}.  However, here we focus only on the CDFs at $z=5.8$ where the two models have the same neutral fraction and comparable likelihoods (i.e. $\mathscr{L}_{z=5.8}$; see Sec. \ref{subsec:step}).}. 

From the figure we see that the difference in the CDFs, {\it when normalized to have the same Lyman alpha transmission}, is modest -- smaller than the cosmic variance of the current sample (see the errorbar for the $50h^{-1}$cMpc result in {\it Ref/BrightGalaxies}).
This is due to the fact that the dynamic range probed by the forest is too small to discriminate against different EoR morphologies at a fixed neutral fraction.  Indeed, we confirm that the largest differences between the two models, when averaged over $10h^{-1}$cMpc, occurs at $\tau_{\rm eff}{\sim}90$.  This is far beyond the observable range.  In matching the same mean flux, the differences in $\tau_{\rm eff}$ appear in the tail of the PDF, and are thus difficult to constrain.
A larger quasar sample ($N_{\rm sample}{\sim}1000$) would narrow down the cosmic variance uncertainties within the accessible range; however,  such a large sample might only become feasible with next-generation telescopes such as {\it Euclid}.

Therefore, we conclude that the current {\lya} forest data is unlikely to be able to distinguish between different galaxy models having the same global properties (such as the EoR history; see also \citealt{Nasir2020MNRAS.494.3080N} for a similar conclusion). In future work, we will also forward model the Lyman-$\beta$ forest, quantifying if its added dynamic range can further help constrain models (as has been implied by, e.g.~\citealt{Eilers2019ApJ...881...23E, Keating2020MNRAS.497..906K}), despite the added cosmic variance from the overlapping, lower redshift {\lya} forest.  We will also investigate the constraining power of other summary statistics.

\subsection{Comparison to previous works}\label{subsec:discussion} 

\begin{figure}
	\begin{minipage}{\columnwidth}
		\includegraphics[width=\textwidth]{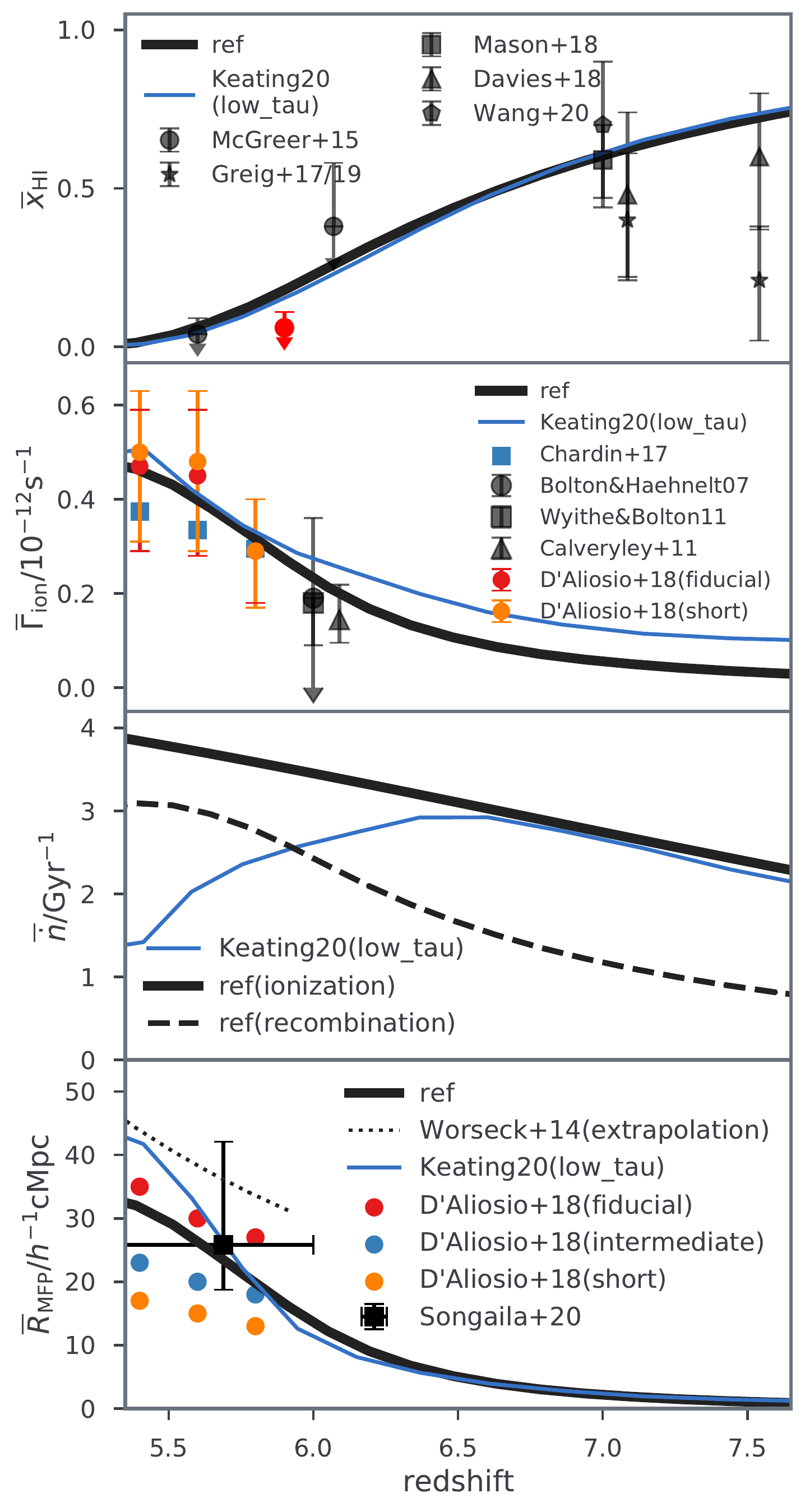} 
	\end{minipage}
	\caption{\label{fig:ML_MFP}The average neutral hydrogen fraction, photoionization rate, ionizing emissivity/recombination rate, mean free path and gas temperature of our maximum likelihood parameter combination ({\it ref}), along with some other models \citep{Chardin2017MNRAS.465.3429C,DAloisio2018MNRAS.473..560D,Keating2020MNRAS.497..906K} and observational limits \citep{Bolton2007MNRAS.382..325B,Wyithe2011MNRAS.412.1926W,Calverley2011MNRAS.412.2543C,Worseck2014MNRAS.445.1745W,McGreer2015MNRAS.447..499M,Greig2017MNRAS.466.4239G,Greig2019MNRAS.484.5094G,Davies2018ApJ...864..142D,Mason2018ApJ...856....2M,Wang2020ApJ...896...23W}}
\end{figure}

Previous studies have made significant progress in understanding the large-scale {\lya} opacity fluctuations at $z\gsim 5.5$. However, since forest simulations are generally computationally expensive, few studies do exhaustive parameter exploration.  Instead, most focus on one or two of the following dominant sources of fluctuations -- the UVB (either through rare sources or a short mean free path), the gas temperature, and late reionization.  In this section we compare our maximum likelihood model from the {\it forest} MCMC (labeled {\it ref}) to some of the recent works that managed to explain the large-scale opacity fluctuations. In Fig.~\ref{fig:ML_MFP} we show the average EoR history, photoionization rate, ionizing emissivity/recombination rate and mean free path, from {\it ref}\footnote{In principle, our Bayesian framework allows us to compare the posterior distributions of these quantities.  However, we do not output all of these fields when running the MCMC. Therefore, here we only show the maximum likelihood model that was re-run to output the desired quantities.} along with some other models and observational data.

As demonstrated quantitatively in Section \ref{subsec:varying fields}, the patchy UVB is an important source of forest fluctuations.  However, we find that UVB fluctuations alone are not sufficient, with late reionization required to fully explain the data (see also, e.g.~\citealt{Keating2020MNRAS.497..906K,Choudhury2020arXiv200308958C}).
Some previous works have focused on the patchy UVB, sourced by either rare, bright AGN or a small mean free path.  We briefly discuss each in turn.

Although rare, bright sources could explain the $\tau_{\rm eff}$ CDF \citep{Chardin2015MNRAS.453.2943C,Chardin2017MNRAS.465.3429C, DAloisio2017MNRAS.468.4691D, Meiksin2020MNRAS.491.4884M}, they somewhat struggle to reproduce the longest GP trough found in ULAS J0148+0600, and would likely be in tension with IGM temperature measurements (e.g.~\citealt{DAloisio2017MNRAS.468.4691D}).  Furthermore, extrapolations of AGN LFs and semi-analytic models suggest they provide a negligible contribution to the UVB at $z>5.5$ (e.g.~\citealt{Mitra2015,Manti2017,Parsa2017,Qin2017a,Garaldi2019MNRAS.483.5301G}).

On the other hand, a small mean free path could modulate a galaxy-dominated UVB to the level required to explain the forest observations (e.g.~\citealt{Davies2016MNRAS.460.1328D, DAloisio2018MNRAS.473..560D}).
However, these models generally require very short mean free paths (c.f.~the {\it early-reionization-short-mfp} model from \citealt{Nasir2020MNRAS.494.3080N} and the {\it short mean free path} from \citealt{DAloisio2018MNRAS.473..560D}  denoted with the orange points in the bottom panel of Fig.~\ref{fig:ML_MFP}).  As discussed in \citet{DAloisio2018MNRAS.473..560D}, observational estimates of the mean free path (e.g.~\citealt{Songaila2010ApJ...721.1448S,Worseck2014MNRAS.445.1745W}) may be biased high, due to contamination from the proximity zone in which the QSO flux dominates over that of the UVB.  Indeed, the mean free path from our ML model is in good agreement with the fiducial values computed in that work, but significantly above those required to explain the opacity fluctuations without evoking a late reionization.  Unlike these works, our framework does not have a tuning knob for the mean free path, instead calculating this quantity directly from the source and sink distributions following \citet{Sobacchi2014MNRAS.440.1662S}.  Our galaxy models and associated parameter priors do not result in mean free paths of ${<}10$ cMpc at $z<6$.

In addition to a patchy UVB, large-scale temperature fluctuations have been evoked to explain the forest observations 
\citep{DAloisio2015ApJ...813L..38D}. 
However, as discussed in Section \ref{subsec:varying fields}, we find temperature fluctuations have only a minor impact (see also \citealt{Keating2018MNRAS.477.5501K}).  This could be due to the fact that strong temperature fluctuations require fairly extended reionization histories, which are disfavored in our posterior.  Indeed, the fiducial model in \citet{DAloisio2015ApJ...813L..38D} is in mild tension with the latest {\it Planck} measurement \citep{Planck2018arXiv180706209P}. Furthermore, these models require that troughs in the transmission come from large scale overdensities, which host an overabundance of galaxies and thus reionize early. This is contrary to recent observations showing a dearth of galaxies around the long GP trough of ULAS J0148+0600 \citep{Becker2018ApJ...863...92B,Kashino2020ApJ...888....6K}.

Finally, late reionization (e.g.~\citealt{Mesinger2010MNRAS.407.1328M}) has also been used to explain the $\tau_{\rm eff}$ distributions \citep{Kulkarni2019MNRAS.485L..24K,Keating2020MNRAS.497..906K,Keating2020MNRAS.491.1736K,Nasir2020MNRAS.494.3080N}. Our results are consistent with this claim (see also the recent work by \citealt{Choudhury2020arXiv200308958C}).  
In Fig.~\ref{fig:ML_MFP} we show the fiducial model from \citet{Keating2020MNRAS.497..906K}, together with our maximum likelihood model.  Amazingly, the EoR histories are in perfect agreement, despite the fact that \citet{Keating2020MNRAS.497..906K} did not perform Bayesian inference.  However, their model requires a drop in the ionizing emissivity ($\overline{\dot{n}}_{\rm ion}$) at $z<6.5$, which is difficult to justify physically and is in contradiction with the observed redshift evolution of the star formation rate density (e.g. \citealt{Bouwens2015a}).
On the other hand, our model achieves the same EoR history without a non-monotonic evolution of the emissivity; the recombination rate increases, approaching the ionization rate, resulting in relatively slow, ``photon starved" end to reionization (e.g.~\citealt{Bolton2007MNRAS.382..325B, Sobacchi2014MNRAS.440.1662S}).
 One explanation for this difference might be that the simulations of \citet{Keating2020MNRAS.497..906K}, with a gas particle mass of $\sim10^7 M_\odot$, could be under-resolving small gas clumps and thus under-estimating the impact of recombinations.  Although the mean free path (being an instantaneous, volume averaged quantity) does not directly show the cumulative impact of inhomogeneous recombinations, the rapid rise in their mean free path below $z<6$ seen in the bottom panel supports this explanation.

\section{Conclusions}\label{sec:conclusions}

In this work, we extend the Bayesian inference framework of {\tocm}/{\cmmc} to forward-model the (low resolution) {\lya} forest. We run MCMCs by sampling empirical galaxy scaling relations and computing the corresponding 3D lightcones of the {\lya} forest.  With these, we quantify the additional constraining power provided by observations of large-scale opacity fluctuations (i.e.~PDFs of $\tau_{\rm eff}$ averaged over 50 cMpc$/h$; \citealt{Bosman2018MNRAS.479.1055B, Bosman2020arXiv200610744B}).  

We find that, in order to be consistent with the observations, our models {\it require} late reionization.  The final overlap stages of the EoR, corresponding to when ${>}95$\% of the volume was ionized, occur at $z{<}5.6$.  Our Bayesian framework provides statistical proof of previous suggestions that the EoR might have completed reionization at $z{<}6$ \citep{Lidz2007ApJ...670...39L,Mesinger2010MNRAS.407.1328M,Kulkarni2019MNRAS.485L..24K,Keating2020MNRAS.491.1736K,Nasir2020MNRAS.494.3080N},
and is perfectly consistent with the recent, similar analysis by \citet{Choudhury2020arXiv200308958C}. Such late reionization is in mild tension ($\sim 1.5 \sigma$) with the dark fraction upper limits from \citep{McGreer2015MNRAS.447..499M}. In the future, we will revisit this mild tension using updated data from the XQR-30 large VLT program.

We also find that the forest data improves our current knowledge of galaxy UV ionizing properties.  In particular, we find a weak ($\sim 1 \sigma$) constraint on the turn-over halo mass scale ($M_{\rm turn}=2{\times}10^{8}$--$10^9\msol$), below which star formation stops being efficient.  Moreover, we find that the late reionization preferred by the forest data tightens constraints on the ionizing escape fraction.  Combined observations (i.e. galaxy UV LFs, CMB optical depth, dark fraction and forest) favor a characteristic ionizing escape fraction of $f_{\rm esc}= 6.9^{+3.8}_{-2.6}$\%, and disfavor a strong evolution with the galaxy's halo (or stellar) mass.  Unfortunately, the $\tau_{\rm eff}$ CDFs cannot distinguish among different source/sink models that have different EoR morphologies but the same EoR history.

Using our maximum-likelihood model, we demonstrate that large-scale opacity fluctuations are driven by a combination of both patchy reionization and spatial variations in the photoionizing background (with temperature inhomogenities being sub-dominant).  The cosmic {$\hone$} patches and regions of weak UVB both corresponds to large-scale underdensities in the matter field.  Thus the longest Gunn-Peterson (GP) troughs correlate with a relative dearth of galaxies, in agreement with observations \citep{Becker2018ApJ...863...92B,Kashino2020ApJ...888....6K} and some previous models (e.g. \citealt{Davies2018ApJ...860..155D}).

Our inference framework can easily be extended to include different source models, such as AGN and/or having a more complex parametrization of galaxy evolution.  Using the Bayesian evidence, we can quantify if the data require the additional model complexity (e.g.~\citealt{Qin2020MNRAS.tmp.3221Q}).
We postpone such investigation to future work, applying them on upcoming, larger data sets.

\section*{Acknowledgements}
We thank George Becker, Anson D'Aloisio, Frederick Davies, Laura Keating, and Girish Kulkarni for helpful comments on a draft version of this paper. This work was supported by the European Research Council (ERC) under the European Union’s 
Horizon 2020 research and innovation programmes (AIDA -- \#638809, First Light -- \#669253 and Cosmic Gas -- \#740246).  The results presented here reflect the authors’ views; the ERC is not responsible for their use.  We acknowledge computational resources of the HPC center at SNS.
MV is supported by grants INFN INDARK PD51 and ASI-INAF n.2017-14-H.0.

\section*{Data availability}
The data underlying this article will be shared on reasonable request to the corresponding author.

\bibliographystyle{\dir mn2e}
\bibliography{reference}

\appendix

\section{Testing the fluctuating Gunn-Peterson approximation}\label{app:sec:FGPA}
\begin{figure*}
	\begin{minipage}{\textwidth}
		\centering
		\includegraphics[width=\textwidth]{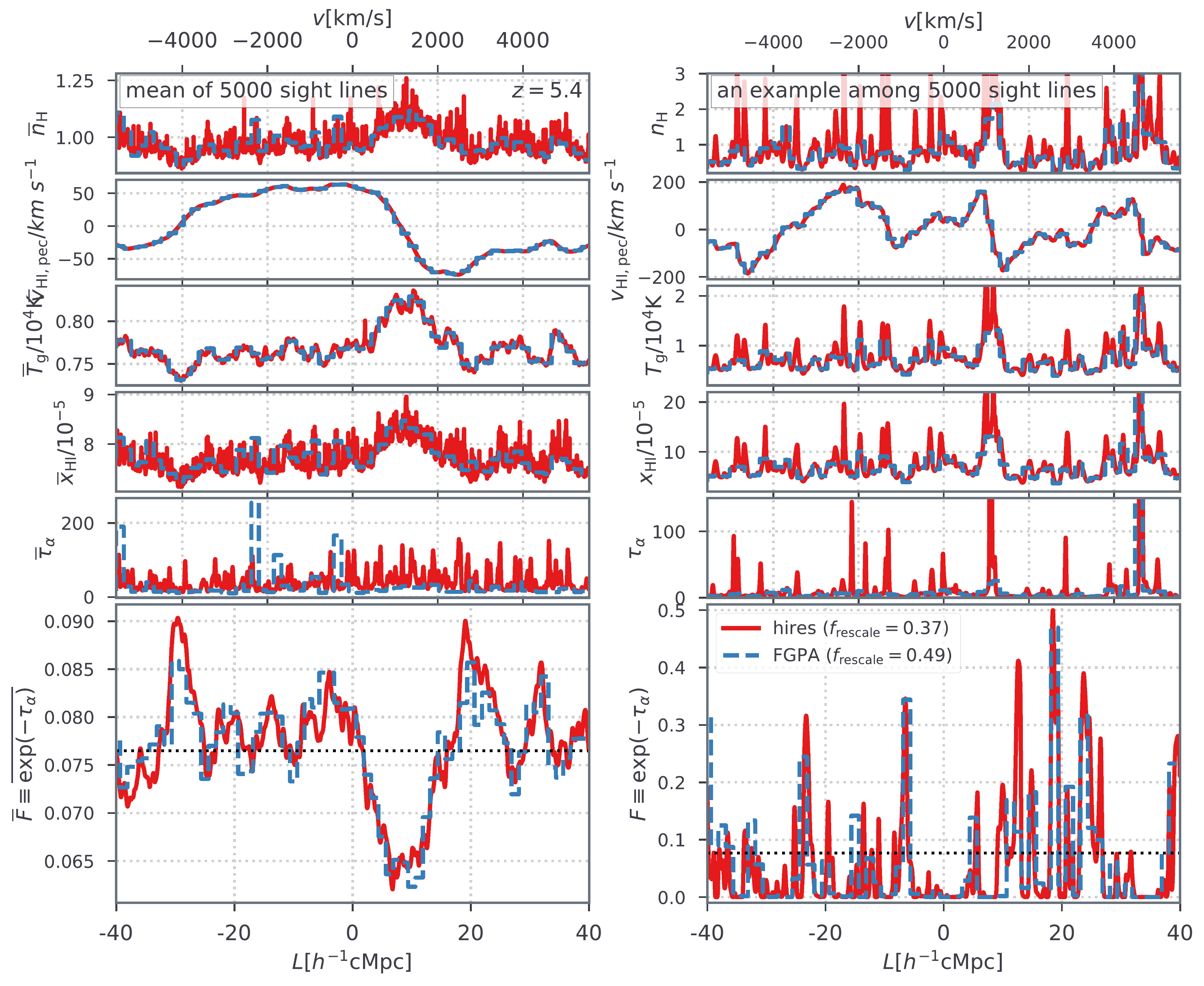}
		\caption{\label{fig:fgpa} Properties including, from top to bottom, gas density, peculiar velocity, gas temperature, residual neutral hydrogen fraction, {\lya} optical depth and its transmission after rescaling the mean flux to the observed one (indicated by the horizontal dotted lines; \citealt{Bosman2020arXiv200610744B}), along 5000 lines of sight across the entire Sherwood simulation box ($80h^{-1}$cMpc) at $z=5.4$. Note that the mean of all sightlines are presented on the left while the right ones show one example sightline. The red solid curves indicate results from the reference model having high spectral resolution and integrating over the full {\lya} profile when evaluating its optical depth ({\it hires}). The blue dashed lines correspond to results assuming FGPA and a lower resolution similar to what is used in the main context ({\it FGPA}). }
	\end{minipage}
\end{figure*}

By computational necessity, our forward models of the Lyman alpha forest are low resolution (${\sim} 2$cMpc cells), and use the fluctuating Gunn-Peterson approximation (FGPA).  The lack of small-scale structure in our models could also impact the large-scale opacity fluctuations we use as our summary statistic (see, e.g. \citealt{Viel2006MNRAS.365..231V} for an example of the impact on the flux power spectrum at lower redshifts where the transmission is higher).  In order to account for this source of inaccuracy, here we compute an error covariance matrix, using a high-resolution hydro simulation from the Sherwood suite \citep{Bolton2017MNRAS.464..897B}.

The simulation used in this section was run with an updated version of \textsc{Gadget}-2 \citep{Springel2005Natur.435..629S} and assumes a $\Lambda$CDM cosmology with parameters ($\Omega_{\mathrm{m}}, \Omega_{\mathrm{b}}, \Omega_{\mathrm{\Lambda}}, h, \sigma_8, n_s $ = 0.31, 0.048, 0.69, 0.68, 0.83, 0.96) from \citet{Planck2016A&A...594A..13P}.  It 
includes $512^3$ baryonic and $512^3$ dark matter particles within a cube of $80h^{-1}$cMpc on a side.  The forest is calculated on a 2048$^3$ Eulerian grid, with a corresponding resolution of 0.057 cMpc (i.e. $80h^{-1}{\rm cMpc}/2048$).

We compute Ly$\alpha$ spectra and $\tau_{\rm eff}$ PDFs (i) assuming the FGPA on fields smoothed down to ${\sim}2$ cMpc resolution; and (ii) using the native high-resolution fields (0.057 cMpc), including peculiar velocities, and a Voigt profile for the Ly$\alpha$ absorption.  The former corresponds to approximations used in our forward-models, while the latter we take as the ``true'' spectra.

We present the results in Fig. \ref{fig:fgpa}, including the hydrogen density, peculiar velocity, gas temperature, residual neutral hydrogen fraction, and the inferred {\lya} optical depth as well as its transmission after rescaling the mean flux to the observed one (i.e. 0.0765; \citealt{Bosman2020arXiv200610744B}).  Since the hydro simulation does not include patchy reionization, we focus on the $z=5.4$ snapshot (this is at the lowest end of the redshift range of interest, where patchy reionization should have the smallest impact; see Fig. \ref{fig:varying_CDF}).

As expected, having a lower resolution reduces the small-scale structure in the forest.  Compared to the high-resolution forest, the FGPA has a smaller variance on small-scales and a $\sim30$\% larger $f_{\rm rescale}$ is required to match the same mean flux.

\begin{figure}
	\begin{minipage}{\columnwidth}
		\centering
		\includegraphics[width=\textwidth]{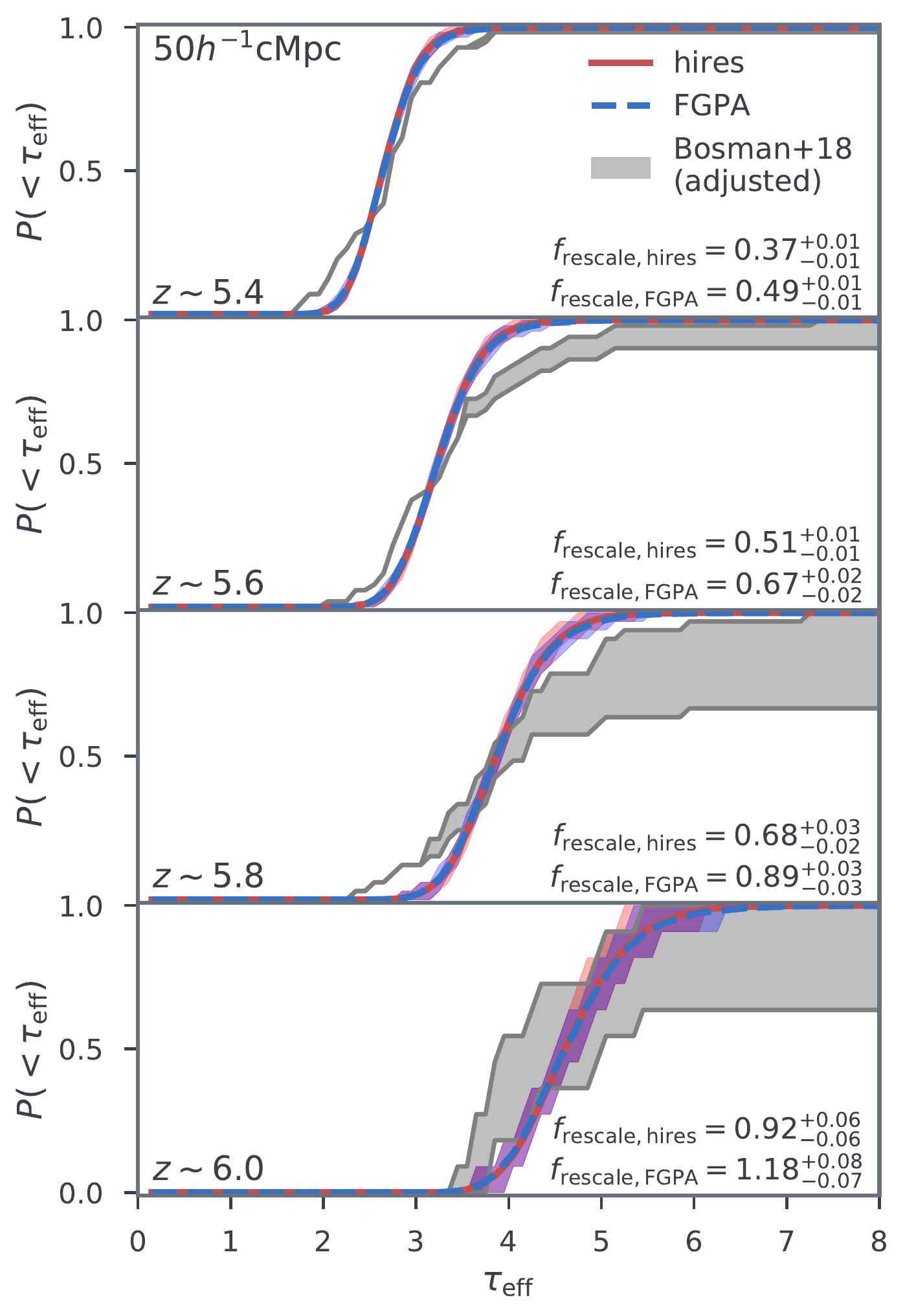}\vspace*{-2mm}
		\caption{\label{fig:fgpa_cdf}  The CDF of $\tau_{\rm eff}$ (averaged over $50h^{-1}$cMpc) at $z=5.4$--6 from the high-resolution Sherwood hydrodynamical simulation ({\it hires}; red) and assuming FGPA with low resolution ({\it FGPA}; blue). The colored lines with shaded region indicate the mean and [14,86] percentiles drawn from 500 realizations. The observed CDFs are indicated in grey (see more in Section \ref{sec:observation}).}
	\end{minipage}
\end{figure}

However, the differences are much smaller in the $\tau_{\rm eff}$ CDFs, averaged over $50h^{-1}$ cMpc.  These are shown in the top panel of Fig. \ref{fig:fgpa_cdf} for both the high-resolution spectra and the FGPA.  We include also the observational data in gray.  As expected, the simulation cannot match the observed distributions, owing to its small size and uniform UVB.  However, the fact that the FGPA and the high-resolution spectra can produce comparable large-scale opacity fluctuations is highly encouraging of our approach.

In Appendix \ref{app:sec:ECM} we show the corresponding error covariance matrix.  This error is added to the cosmic variance error used when computing the forest likelihood in our MCMCs.

\section{Error Covariance Matrices}\label{app:sec:ECM}
\begin{figure}
	\begin{minipage}{\columnwidth}
		\centering
		\includegraphics[width=\textwidth]{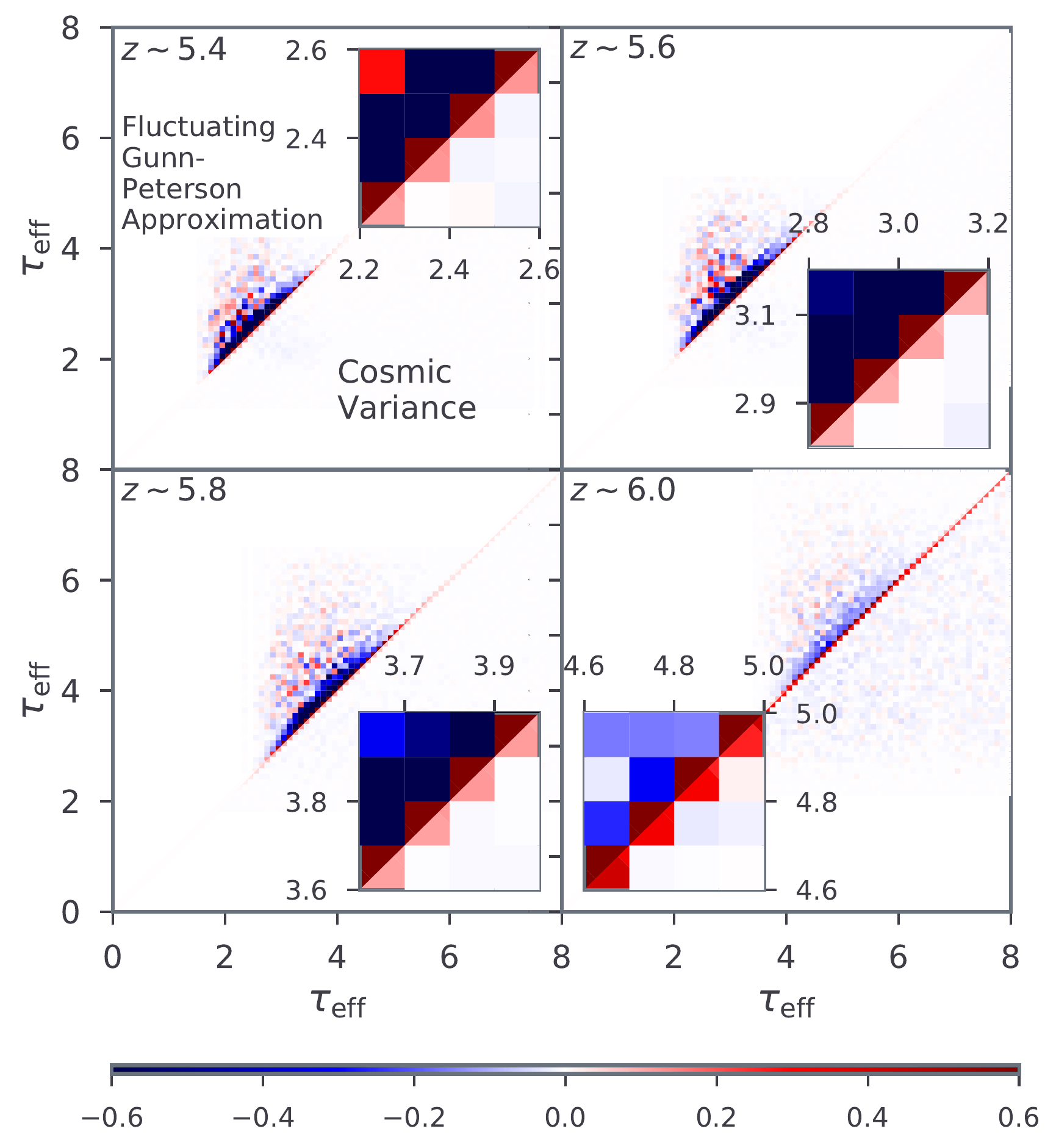}
		\caption{\label{fig:ecm} The error covariance matrices (ECMs) of the $\tau_{\rm eff}$ PDF introduced by the fluctuating Gunn-Peterson approximation (FGPA) and cosmic variance (CV) at $z=5-6$. The two symmetric matrices are presented together in each sub-panel separated by the diagonal. The coefficients of the ECMs are represented by varying colors shown in the colorbar. A zoom-in sub-panel is presented to show more details around the maximum diagonal element of either matrix.}
	\end{minipage}
\end{figure}

We present the error covariance matrix\footnote{An identity matrix with a normalization of $10^{-5}$ is imposed on the total error covariance matrix as a precautionary measure.} (ECM) of the $\tau_{\rm eff}$ PDF introduced by making the fluctuating Gunn-Peterson approximation (FGPA) or from the cosmic variance (CV) at $z=5.4-6$ in Fig. \ref{fig:ecm}. 
Since ECMs are symmetric, we only show half of the FGPA and CV matrices, and present them together (i.e. top is FGPA and bottom is CV, respectively). 
For this plot, the CV is estimated using the 500 realizations\footnote{For computing efficiency, only 150 realizations are generated during the MCMC.} drawn from our ML model (see more in Section \ref{sec:models}).  We note that the cosmic variance ECM is re-computed on the fly for each sample in our MCMC; unfortunately,  it is computationally impractical to do this also for the FGPA ECM \footnote{As we do not know a-priory what are the true values of these fields in the real Universe, the FGPA error covariance matrix should be recomputed for each forward modelled universe. In other words, we use $\Sigma_{\rm GP}(\theta = \theta_{\rm Sherwood})$ in our MCMC, where $\theta_{\rm Sherwood}$ encapsulates all of the choices and approximations made to generate the Sherwood simulation. However, we should instead know the general error covariance, $\Sigma_{\rm GP}(\theta)$, evaluating it on-the-fly for any astrophysical parameter combination $\theta$. Unfortunately, this is computationally impractical. Using different hydro simulations, in the future we will explore how the covariance matrix changes for a few different values of $\theta$, and estimate the corresponding impact on the posteriors.}.
 The latter was computed as described in the previous Appendix at $z=5.4$, and adjusted for higher redshifts by shifting the PDF such that the mean flux matches the observations (see the corresponding CDFs in Fig. \ref{fig:fgpa_cdf}).  

There are both positive (correlation; red) and negative values (anti-correlation; blue) between pairs of different $\tau_{\rm eff}$ bins, with more showing anti-correlation when being closer to the diagonal (i.e. nearby bins).  This is expected as increasing one histogram bin can be roughly compensated by a decrease in a nearby bin.  The dominant component in the total ECM is caused by the diagonal coefficients of the FGPA, though the  and its coefficients along the diagonal (i.e. uncertainties in each $\tau_{\rm eff}$ bin of the PDF) decreases towards lower and higher redshifts. This is because when the sample size decreases, the PDF, with a lower value, also possesses a smaller absolute difference between the FGPA and the reference results (e.g. see the corresponding CDF in the lower right panel of Fig. \ref{fig:fgpa}). On the other hand, the CV ECM shows the opposite trend, which is also caused by the sample size. Since different lines of sight are randomly selected when estimating the cosmic variance, a smaller sample size leads to larger differences between different realizations and therefore larger CV.

\label{lastpage}
\end{document}